\documentclass[prd,nofootinbib,a4paper]{revtex4}
\usepackage[a4paper, hdivide={2.5cm,,2.5cm}, vdivide={2.5cm,,2.5cm}]{geometry}
\pdfoutput=1
\usepackage{amstext,amssymb}
\usepackage[intlimits]{amsmath}
\usepackage[english]{babel}
\usepackage[ansinew]{inputenc}
\usepackage{graphicx}
\usepackage{slashed} 
\usepackage{units}
\usepackage{enumerate}
\usepackage{hyperref}

\hypersetup{
    pdftitle={Gauged \texorpdfstring{$L_\mu - L_\tau$}{L(mu)-L(tau)} Symmetry at the Electroweak Scale},
    pdfauthor={Julian Heeck, Werner Rodejohann}
}

\pacs{14.70.Pw, 12.60.Cn, 12.60.Fr, 14.60.Pq, 14.60.St}

\begin{document}

\let \Lold \L
\def \L {\mathcal{L}} 
\let \epsilonold \epsilon
\def \epsilon {\varepsilon} 
\let \arrowvec \vec
\def \vec#1{{\boldsymbol{#1}}}
\def\ra{\rightarrow}
\newcommand{\Uprime}{U(1)_{L_\mu-L_\tau}}
\newcommand{\del}{\partial}
\newcommand{\dd}{\mathrm{d}}
\newcommand{\matrixx}[1]{\begin{pmatrix} #1 \end{pmatrix}} 
\newcommand{\tr}{\mathrm{tr}}
\newcommand{\hc}{\mathrm{h.c.}}
\newcommand{\sm}{\mathrm{SM}}
\newcommand{\BR}{\mathrm{BR}}
\newcommand{\re}{\mathrm{Re}\,}
\newcommand{\im}{\mathrm{Im}\,}
\newcommand{\dof}{\mathrm{d.o.f.}}

\title{Gauged $\boldsymbol{L_\mu - L_\tau}$ Symmetry at the Electroweak Scale}

\author{Julian \surname{Heeck}}
\email{julian.heeck@mpi-hd.mpg.de}
\affiliation{Max--Planck--Institut f\"ur Kernphysik,\\Postfach 103980, D--69029 Heidelberg, Germany}

\author{Werner \surname{Rodejohann}}
\email{werner.rodejohann@mpi-hd.mpg.de}
\affiliation{Max--Planck--Institut f\"ur Kernphysik,\\Postfach 103980, D--69029 Heidelberg, Germany}

\begin{abstract}
The extension of the Standard Model by a spontaneously broken abelian
gauge group based on the $L_\mu - L_\tau$ lepton number 
can resolve the long-standing discrepancy between 
experimental and theoretical values for the magnetic moment of the
muon. It furthermore naturally generates $\mu$-$\tau$ symmetric
lepton mixing, introduces neutrino nonstandard interactions, 
and the associated gauge boson $Z'$ serves as a
mediator to the right-handed neutrino sector. 
A detailed fit to electroweak data is performed to identify the
allowed values for the mass of $Z'$ and its mixing with the Standard
Model $Z$. An economical new scalar sector is constructed that
spontaneously breaks $L_\mu - L_\tau$ and leads to experimental 
consequences such as lepton flavor violation and collider signatures. 
Furthermore we discuss the nonabelian extension to an
$SU(2)'$, particularly the neutrino sector.
\end{abstract}

\maketitle

\section{Introduction}

Extensions of the Standard Model (SM) of particle physics could either add 
new particles or representations, or extend the gauge sector. The
enormous precision with which the SM has been tested in the last
decades, plus the various theoretical consistency conditions which have to be
obeyed, require careful addition of new physics. 
One particularly popular approach is the addition of an abelian gauge
symmetry $U(1)'$. If this symmetry is broken (to avoid 
an additional force with infinite range), a massive $Z'$ boson is present, with
model-dependent mass and couplings to the SM particles
\cite{PL_review}. In this paper we focus on one class of highly
interesting $U(1)'$ models: within the particle content of the SM, it
is possible to gauge one of the three differences of lepton flavors 
$L_e - L_\mu$, $L_e - L_\tau$ or $L_\mu - L_\tau$, without introducing
an anomaly \cite{zero,joshi}. This surprising feature has lead to a
number of works analyzing the consequences of one of those broken symmetries 
\cite{longrange,Choubey:2004hn,Ma,Baek,Gninenko:2001hx}. In particular, $L_\mu -
L_\tau$ should be preferred over the other two combinations, because
in the limit of conserved symmetry, the neutrino mass matrix is
automatically $\mu$-$\tau$ symmetric, and predicts one degenerate
neutrino pair. The necessary breaking of the symmetry will split their
masses and generate small departures from $\mu$-$\tau$ symmetry,
thereby rendering the neutrino phenomenology in agreement with data. In
contrast, if $L_e - L_\mu$ or $L_e - L_\tau$ are to be gauged, the
neutrino mass matrix has in the symmetry limit a structure far away
from the one necessary to reproduce the experimental results. 
This intimate connection of flavor and gauge symmetry is rather
unique. It is worth stressing that gauged $L_\mu - L_\tau$ gives a lepton mixing
structure close to observation without the usual complications of
flavor symmetries (see \cite{Altarelli:2010gt} for recent reviews), in
which one typically involves nonrenormalizable terms including a 
plethora of ``flavon fields'', arranges for their proper vacuum expectation value (VEV) alignment
by additional input, and adds additional symmetries to avoid
unwanted terms in the Lagrangian.

An important property of gauged $L_\mu - L_\tau$ is that it 
does not act on first-generation leptons, but only on muons and
tauons. In this respect, further motivation for this model (and the main focus of previous
discussions of this gauge group~\cite{Ma,Baek}) stems from the
anomalous magnetic moment of the muon. This measured quantity 
exhibits a $3.2\sigma$ difference to the theoretically predicted
value, which can be explained by the loop-contribution of a 
heavy $Z'_{L_\mu-L_\tau}$ gauge boson.

In the present work we study the phenomenology of gauged $L_\mu -
L_\tau$ in the regime in which the anomalous magnetic moment of the
muon is explained. In Section~\ref{sec:gauge_sector} we determine the
currently allowed parameter space for a generic $\Uprime$ as
determined by electroweak precision data, and also comment on collider
physics aspects of heavy $Z'$ bosons. We propose a new and economic 
scalar sector to break the symmetry spontaneously, and study the
resulting neutrino sector in Section~\ref{sec:neutrino_sector}. 
The scalar potential and the Higgs spectrum is analyzed in Section~\ref{sec:scalar_sector}. 
Section~\ref{sec:nonabelianextension} is devoted to an extension from
$\Uprime$ to an $SU(2)'$ that also acts on the electron. 
An overview over the used notation concerning nontrivial gauge group representations is
delegated to Appendix~\ref{app:fieldtrafos}, 
while Appendix~\ref{app:reduciblerep} briefly discusses 
leptons in reducible representations of $SU(2)'$. Finally, we 
conclude in Section~\ref{sec:conclusion}. 

\section{Gauge Sector}\label{sec:gauge_sector}
 
Extending the gauge group of the SM $G_\sm \equiv SU(3)_C \times SU(2)_L \times U(1)_Y$ by $\Uprime$ leads to possible $Z$--$Z'$ mixing, even without a scalar charged under both $U(1)$ groups~\cite{PL_review}. This is due to kinetic mixing, i.e.~a gauge-invariant term $\sim Z^{\mu\nu} Z'_{\mu\nu}$ in the Lagrange density $\L = \L_\mathrm{SM} + \L_{Z'} + \L_\mathrm{mix}$, with $Z^{\mu\nu}$ being the gauge field strength tensor. If a scalar transforms nontrivially under $SU(2)_L \times U(1)_Y \times \Uprime$ and acquires a VEV, a mass-mixing term $\sim Z^\mu Z'_\mu$ can also be generated, so the most general Lagrangian after breaking $SU(2)_L \times U(1)_Y \times \Uprime$ to $U(1)_\mathrm{EM}$ takes the form:
\begin{align}
\begin{split}
	\L_\mathrm{SM} &= -\frac{1}{4} \hat{B}_{\mu\nu} \hat{B}^{\mu\nu} -\frac{1}{4} \hat{W}^a_{\mu\nu} \hat{W}^{a\mu\nu} + \frac{1}{2} \hat{M}_Z^2 \hat{Z}_\mu \hat{Z}^\mu  - \frac{\hat{e}}{\hat{c}_W} j_Y^\mu \hat{B}_\mu -\frac{\hat{e}}{\hat{s}_W} j_W^{a\mu} \hat{W}^a_\mu\,,	\\
	\L_{Z'} &= -\frac{1}{4} \hat{Z}'_{\mu\nu} \hat{Z}'^{\mu\nu}+ \frac{1}{2} \hat{M}_Z'^2 \hat{Z}'_\mu \hat{Z}'^\mu - \hat{g}' j'^\mu Z'_\mu \,,\\
	\L_\mathrm{mix} &= -\frac{\sin \chi}{2} \hat{Z}'^{\mu\nu}\hat{B}_{\mu\nu} + \delta \hat{M}^2 \hat{Z}'_\mu \hat{Z}^\mu \,.
\end{split}
\label{eq:lagrangian}
\end{align}
The currents are defined as
\begin{align}
\begin{split}
	j_Y^\mu &= -\sum_{\ell = e,\mu,\tau} \left[\overline{L}_\ell \gamma^\mu L_\ell + 2\, \overline{\ell}_R \gamma^\mu \ell_R \right] + \frac{1}{3} \, \sum_{\mathrm{quarks}} \left[\overline{Q}_L \gamma^\mu Q_L + 4 \,\overline{u}_R \gamma^\mu u_R -2\, \overline{d}_R \gamma^\mu d_R\right]\,,\\
	 j_W^{a\mu} &= \sum_{\ell = e,\mu,\tau}\overline{L}_\ell \gamma^\mu \frac{\sigma^a}{2} L_\ell + \sum_{\mathrm{quarks}}  \overline{Q}_L \gamma^\mu \frac{\sigma^a}{2} Q_L \,,\\
	j'^\mu &= \bar{\mu} \gamma^\mu \mu + \bar{\nu}_\mu \gamma^\mu P_L \nu_\mu - \bar{\tau} \gamma^\mu \tau - \bar{\nu}_\tau \gamma^\mu P_L \nu_\tau \,,
\end{split}
\end{align}
with the left-handed $SU(2)$-doublets $Q_L$ and $L_\ell$ and the Pauli matrices $\sigma^a$. We also define the electric current $j_\mathrm{EM} \equiv j_W^{3} + \frac{1}{2}\, j_Y$ and the weak neutral current $j_\mathrm{NC} \equiv 2 j_W^3 - 2 \hat{s}_W^2 j_\mathrm{EM}$.
We adopt the notation of Ref.~\cite{Babu:1997st} with gauge-eigenstates $\{\hat A, \hat Z,\hat Z'\}$ connected to the mass-eigenstates $\{A,Z_1,Z_2\}$ via:
\begin{align}
	\matrixx{\hat A\\ \hat Z \\ \hat Z'} = \matrixx{1 & -\hat c_W \sin \xi \tan \chi & -\hat c_W \cos\xi \tan \chi\\ 0 & \cos \xi + \hat s_W \sin \xi \tan \chi & \hat s_W \cos \xi \tan \chi - \sin \xi\\ 0 & \frac{\sin\xi}{\cos\chi}& \frac{\cos\xi}{\cos\chi}} \matrixx{A\\ Z_1 \\ Z_2} , 
	\label{eq:mass_eigenstates}
\end{align}
or, inverted:
\begin{align}
 \matrixx{A\\ Z_1 \\ Z_2} = \matrixx{1 & 0 & \hat c_W \sin\chi\\ 0 & \cos\xi & -\hat s_W \cos\xi \sin\chi + \sin\xi \cos\chi\\ 0 & -\sin\xi & \cos\xi \cos\chi + \hat s_W \sin \xi \sin\chi} \matrixx{\hat A\\ \hat Z \\ \hat Z'} .
\end{align}
Here the mixing angle $\xi$ is defined as $\tan 2\xi = \frac{2b}{a-c}$ with
\begin{align}
\begin{split}
	a &\equiv \hat{M}_Z^2 \,,  \qquad b \equiv \hat{s}_W \tan \chi \hat{M}_Z^2 + \frac{\delta \hat{M}^2}{\cos \chi} \,,\\
	c &\equiv \frac{1}{\cos^2 \chi} \left( \hat{M}_Z^2 \hat{s}^2_W \sin^2 \chi + 2 \hat{s}_W \sin \chi \delta \hat{M}^2 + \hat{M}_{Z'}^2\right) \,.
\end{split}
	\label{eq:abc}
\end{align}
The gauge boson couplings to fermions are hence changed to
\begin{align}
\begin{split}
&e j_\mathrm{EM} \hat{A} + \frac{e}{2 s_W c_W} j_\mathrm{NC} \hat{Z}  + g' j' \hat{Z}' \ra \\
&\matrixx{e j_\mathrm{EM}, & \frac{e}{2 \hat s_W \hat c_W} j_\mathrm{NC}, & g' j'}\matrixx{1 & -\hat c_W \sin \xi \tan \chi & -\hat c_W \cos\xi \tan \chi\\ 0 & \cos \xi + \hat s_W \sin \xi \tan \chi & \hat s_W \cos \xi \tan \chi - \sin \xi\\ 0 & \frac{\sin\xi}{\cos\chi}& \frac{\cos\xi}{\cos\chi}} \matrixx{A\\ Z_1 \\ Z_2} .
\end{split}
\label{eq:current_vector_interaction}
\end{align}
In the following we will for simplicity set $\chi = 0$, as the mass
mixing already shows all qualitative effects of mixing and will be
induced in our specific model in Section \ref{sec:scalar_sector}. A nonzero $\chi$ results in an additional coupling of $Z_2$ to the electromagnetic current which will change the constraints on $\sin\xi$ given below~\cite{PL_review}.

The mass eigenstate $Z_1$ was studied extensively at LEP, so we know
its axial and vector couplings to leptons to a high
precision.\footnote{The axial and vector couplings can be obtained by
rewriting $j^\mu_X X_\mu =\sum_\psi \overline{\psi}\, \gamma^\mu
(g_V^\psi - g_A^\psi \gamma^5) \psi\,X_\mu$.} Compared to the SM case,
the $Z_1$ couplings become nonuniversal due to the small admixture of
the $j'$ current. For a small mixing angle $\xi$ this additional
coupling to $g' \xi\, j'^\mu$ modifies, for example, the well-measured
asymmetry parameter $A^\ell = 2 g^\ell_V g^\ell_A /((g^\ell_V)^2 +
(g^\ell_A)^2)$ for muons and tauons:
\begin{align}
	A^\mu \ra A^\mu \,\left( 1- g' \xi \frac{4 s_W c_W/e}{1-4 s_W^2}\right)\,, &&
	A^\tau \ra A^\tau \,\left( 1+ g' \xi \frac{4 s_W c_W/e}{1-4 s_W^2}\right)\,.
\end{align}
Depending on the sign of $g'\xi$, we expect a hierarchy $A^\mu < A^e < A^\tau$ or $A^\tau < A^e < A^\mu$ (the SM predicts $A^e = A^\mu = A^\tau$), neither of which is observed~\cite{PDG2010}. This can be used to estimate a $3\sigma$ constraint $|g' \xi | \lesssim  10^{-3}$.

\begin{figure}[t]
\centering
\includegraphics[scale=0.6]{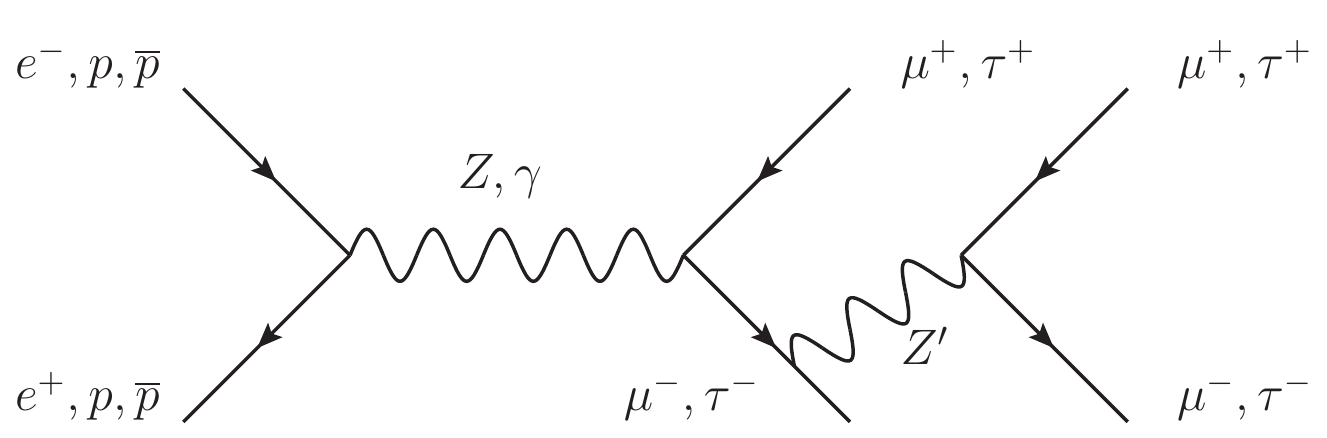}
    \caption{Detection process for an unmixed $Z'$ in electron-positron (LEP) or proton-(anti-)proton (LHC and Tevatron) collisions.}
    \label{fig:ppto4muon}
\end{figure}

In the unmixed case ($\chi=0=\delta \hat{M}^2$), the main constraint on the model stems from the anomalous magnetic moment of the muon, since the $Z'$ contributes a term~\cite{zprimecontribution}
\begin{align}
\begin{split}
	\Delta a_\mu &=\frac{g'^2 }{4\pi} \frac{1}{2\pi} \int_0^1 \dd x\ \frac{2 m_\mu^2 x^2 (1-x)}{m_\mu^2 x^2 + M_{Z'}^2 (1-x)}\\
	&\simeq \frac{g'^2 }{4\pi} \frac{1}{2\pi}
	\begin{cases}
		1 & \mathrm{ for }\  M_{Z'} \ll m_\mu \,,\\
		{2 m_\mu^2}/{3 M_{Z'}^2} &  \mathrm{ for }\  M_{Z'} \gg m_\mu \,,
	\end{cases} 
\end{split}
\label{eq:deltaamu}
\end{align}
which can be used to soften the longstanding $3.2\sigma$ disagreement between experiment ($a_\mu^\mathrm{exp} = 11 659 2089\times 10^{-11}$) and theoretic predictions ($a_\mu^\mathrm{SM} = 116591834\times 10^{-11}$)~\cite{Ma,PDG2010} (note however the uncertainty in the hadronic contributions to $a_\mu$~\cite{hadronic_contrib}). The appropriate values for the case $M_{Z'}\gg m_\mu$ lie around $M_{Z'}/g' \sim \unit[200]{GeV}$:
\begin{align}
	\Delta a_\mu \simeq 236 \times 10^{-11}\, \left( \frac{\unit[200]{GeV}}{M_{Z'}/g'}\right)^2\,.
\end{align} 
The authors of Ref.~\cite{Ma} also derive a direct detection limit of roughly $M_{Z'} >\unit[50]{GeV}$, based on ALEPH measurements of the process $e^+ e^- \rightarrow \mu\mu Z'^* \rightarrow 4\mu$~\cite{aleph} (see Fig.~\ref{fig:ppto4muon}) and under the assumption $M_{Z'}/g' \sim \unit[200]{GeV}$. A similar analysis using LEP2 data with higher luminosity $\sim \unit[0.7]{fb^{-1}}$~\cite{delphi} gives a bound of the same order due to the low rate of $4 \ell$ final states at LEP2 energies. We will comment on Tevatron and LHC prospects of this process in Sec.~\ref{sec:lhc}.

Interestingly, the nonuniversality can also lead to nonstandard
neutrino interactions (NSIs), which are usually parametrized by the
nonrenormalizable effective Lagrangian 
\begin{align}
	\L^\mathrm{eff}_\mathrm{NSI} = -2 \sqrt{2} G_F \epsilon_{\alpha\beta}^{f P} \left[\bar f \gamma^\mu P f \right] \left[ \bar \nu_\alpha \gamma_\mu P_L \nu_\beta\right] \,.
	\label{eq:neutralNSI}
\end{align}
Note that this Lagrangian, when written in a gauge-invariant way,
introduces charged-lepton flavor violation, which usually is a big
problem in phenomenological studies of NSIs. 
Effective four-fermion interactions can be obtained from Eq.~\eqref{eq:lagrangian} by integrating out the (heavy) mass eigenstate $Z_2$ after performing the transformation from Eq.~\eqref{eq:current_vector_interaction}. Since the analytical expression for the NSIs are only marginally more complicated with a nonzero $\chi$, we will include it in the next two equations. The effective Lagrangian for $Z_2$ interactions takes the form
\begin{align}
\begin{split}
	\L^\mathrm{eff}_{Z_2} = \frac{-1}{2 M_2^2}\left( g' \frac{\cos\xi}{\cos\chi}\, j'  -e \hat{c}_W \cos\xi \tan \chi \, j_\mathrm{EM}\right.+ \left.\frac{e}{2 \hat{s}_W \hat{c}_W} (\hat{s}_W \cos \xi \tan \chi - \sin \xi) \,j_\mathrm{NC} \right)^2 \,.
\end{split}
\end{align}
Expanding the square we only need the terms linear in $g'$ for the NSIs, because the
others are either diagonal in flavor space and hence do not influence
neutrino oscillations (these are just the SM terms with $g'=0$), or do
not involve $e$, $u$ or $d$ and hence do not couple neutrinos to
``matter" (the terms quadratic in $g'$). Integrating out $Z_1$ gives
similar terms, so after adding up the different contributions to the
effective, Earth-like matter NSI $\epsilon_{\alpha\beta}^{\oplus}
=\epsilon^{e V}_{\alpha\beta}+3\, \epsilon^{u V}_{\alpha\beta}+3\,
\epsilon^{d V}_{\alpha\beta}$, we obtain the two parameters
$\epsilon_{\mu\mu}^{\oplus}$ and
$\epsilon_{\tau\tau}^{\oplus}=-\epsilon_{\mu\mu}^{\oplus}$ with
\begin{align}
	\epsilon_{\mu\mu}^{\oplus} = \frac{-g'}{4\sqrt{2} G_F \cos\chi}\frac{e}{\hat{s}_W \hat{c}_W} \left[  \cos \xi \sin\xi \left(\frac{1}{M_1^2}-\frac{1}{M_2^2}\right) + \hat{s}_W \tan \chi \left(\frac{\sin^2\xi}{M_1^2} + \frac{\cos^2\xi}{M_2^2}\right)\right]  \,.
\label{eq:epsilon_mumu_earth}
\end{align}
The analogous parameter for solar matter $\epsilon^\odot = \epsilon^e
+ 2\, \epsilon^u + \epsilon^d$ vanishes, which shows that in neutral
matter the potential is generated by the neutrons.\footnote{Correspondingly, the kinetic mixing angle $\chi$---describing a coupling to the electromagnetic current---gives only minor contributions to $\epsilon_{\alpha\beta}^{\oplus}$ and can not generate them without $\delta \hat M^2$. This can be seen from Eq.~\eqref{eq:epsilon_mumu_earth} in the limit $\xi,\chi \ll 1$ or, more general, by using the parameters $( \hat M_i,\delta \hat M^2,\chi)$ instead of $( M_i,\xi,\chi)$.} While this is
similar to gauged $B-L$ symmetries, the nonuniversality of our model
could make these NSIs observable in neutrino oscillations.
For instance, taking $\chi=0$, $g' \simeq
1/4$, $M_2 \simeq \unit[50]{GeV}$ and $\xi \simeq 4 \times 10^{-3}$
can generate 
\begin{align}
\epsilon^\oplus_{\mu\mu} \simeq 10^{-2}\textrm{--}10^{-3} \, , 
\end{align} 
which is testable in future facilities \cite{nufact} and can 
and still resolve the magnetic moment of the muon $\Delta a_\mu$. The mixing
angle $\xi$ obviously has to be not too tiny for such an effect to be
observable. In addition, the mass of the $Z'$ should not be too
heavy: for $M_2 \gg M_1$, $\xi \ll 1$ and $\chi=0$ 
Eq.~(\ref{eq:epsilon_mumu_earth}) can be simplified to 
$\epsilon^\oplus_{\mu\mu} \simeq 
\frac{-g'  \xi}{4 \sqrt{2} G_F}\frac{e}{s_W c_W} \,  \frac{1}{M_1^2}
\simeq -1.5\, g' \xi$, where the constraint 
$g' \xi \lesssim 10^{-3}$ by $Z$-pole measurements 
(even stronger limits are given below) suppresses the NSI parameter. 
Since the two gauge boson masses
enter with opposite sign, the NSI parameters will be even smaller in
the limit $M_1 \sim M_2$, only $M_2 < M_1$ can lead to NSI values
closer to the current limit. 

Nevertheless, there is allowed parameter space of the model allowing
for testable NSI, providing a  
complementary way to probe such nonuniversal gauge bosons. Note
further that this renormalizable realization of NSI parameters does, in contrast 
to the effective approaches as in Eq.~(\ref{eq:neutralNSI}), 
not suffer from charged-lepton flavor violation decays, 
due to the diagonal structure of the NSI parameters
$\epsilon_{\alpha\beta}$ in flavor space, as imposed by our symmetry. 
Consequently, lepton flavor violation will enter into our model only
via the symmetry breaking sector, which we will discuss in Sec.~\ref{sec:scalar_sector}.

\subsection{Fit to Electroweak Precision Data}
\label{sec:gappfit}
Using a recent version (April 2010) of the Fortran program GAPP
(Global Analysis of Particle Properties)~\cite{Erler}, which we
modified to take the $Z'_{L_\mu-L_\tau}$ boson into account, we can
fit the $\Uprime$ model -- with a mass around the electroweak scale -- to
a vast amount of electroweak precision data, including radiative
corrections of the Standard Model. In the following we will set the
kinetic mixing angle $\chi$ to zero, as its inclusion will not alter
the discussed phenomenology qualitatively~\cite{PL_review}. Its effect
would be most interesting for symmetry breaking schemes that do not
generate $\delta \hat M^2$ (e.g.~$U(1)'$ breaking via SM-singlet scalars),
which is not the case for the model discussed in
Section~\ref{sec:scalar_sector}.

We leave the Higgs sector unspecified for the analysis as to not
introduce more parameters, but restrict it to singlets and doublets
under $SU(2)_L$, i.e.~we leave the $\rho$ parameter untouched.  
Since the modified couplings of $Z_1$ compared to the SM only involve
the mixing angle $\xi$ in combination with the coupling constant
$g'$~(see Eq.~\eqref{eq:current_vector_interaction}), while $Z_2$
mainly contributes to $\Delta a_\mu$~(see Eq.~\eqref{eq:deltaamu}), we use the
common convention~\cite{PL_review} of giving limits on the two
quantities $g' \sin\xi$ and $M_2/g'$.

As fit parameters, we used the conventional set of masses ($Z$, Higgs, top-, bottom- and charm-quark) and couplings (strong coupling constant $\alpha_s$ and the radiative contribution of the lightest three quarks to the QED coupling constant $\Delta \alpha_\mathrm{had}^{(3)}$\footnote{Contributing to the on-shell coupling via $\alpha (M_Z) = \alpha/[1-\Delta \alpha (M_Z)]$.}), listed in Table~\ref{tab:gapp_bestfit}. We also enforced the direct $95\%$ C.L.~exclusion limit $m_H > \unit[114.4]{GeV}$ given by LEP; since a Higgs mass $\sim \unit[90]{GeV}$ is favored by an unconstrained fit, $m_H$ lies at its lower bound. 
Except for $M_2$ and $\sin \xi$, we will not bother calculating the errors on the best-fit values, since they are not of interest here. 

\begin{figure}[t]
\centering
\includegraphics[height=0.33 \textwidth]{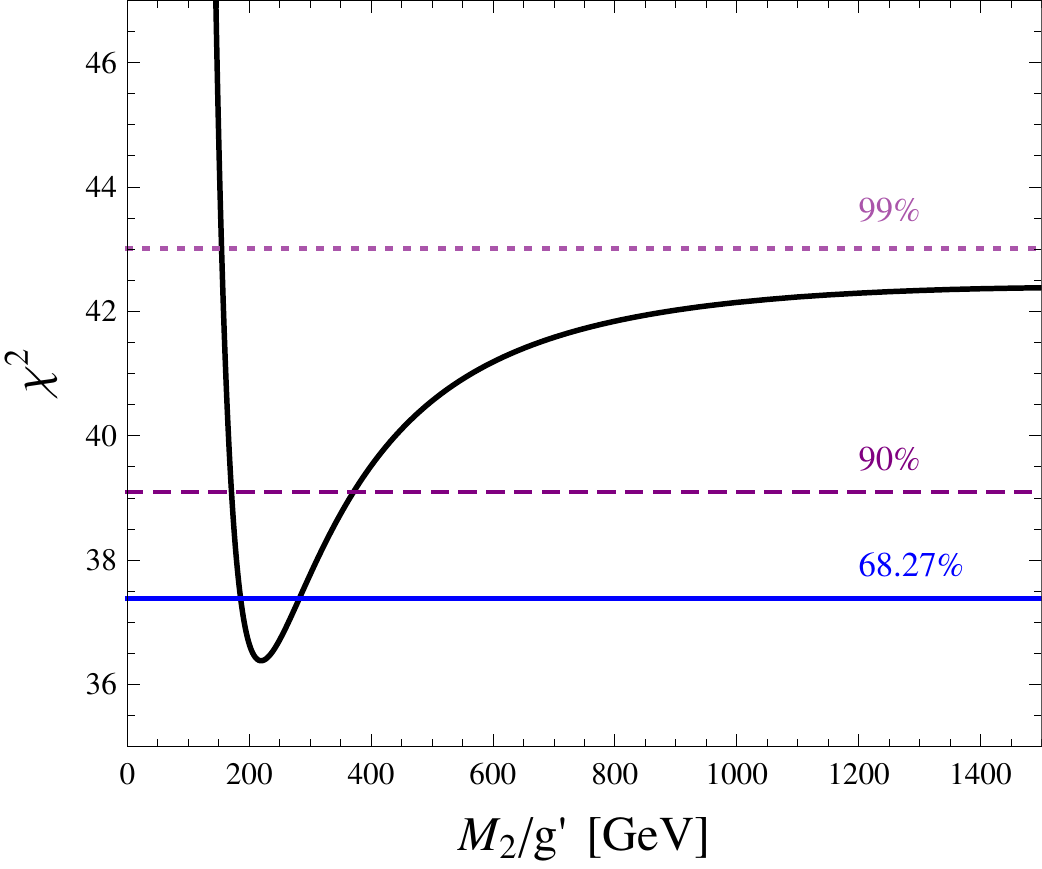}\hspace{2ex}
\includegraphics[height=0.33 \textwidth]{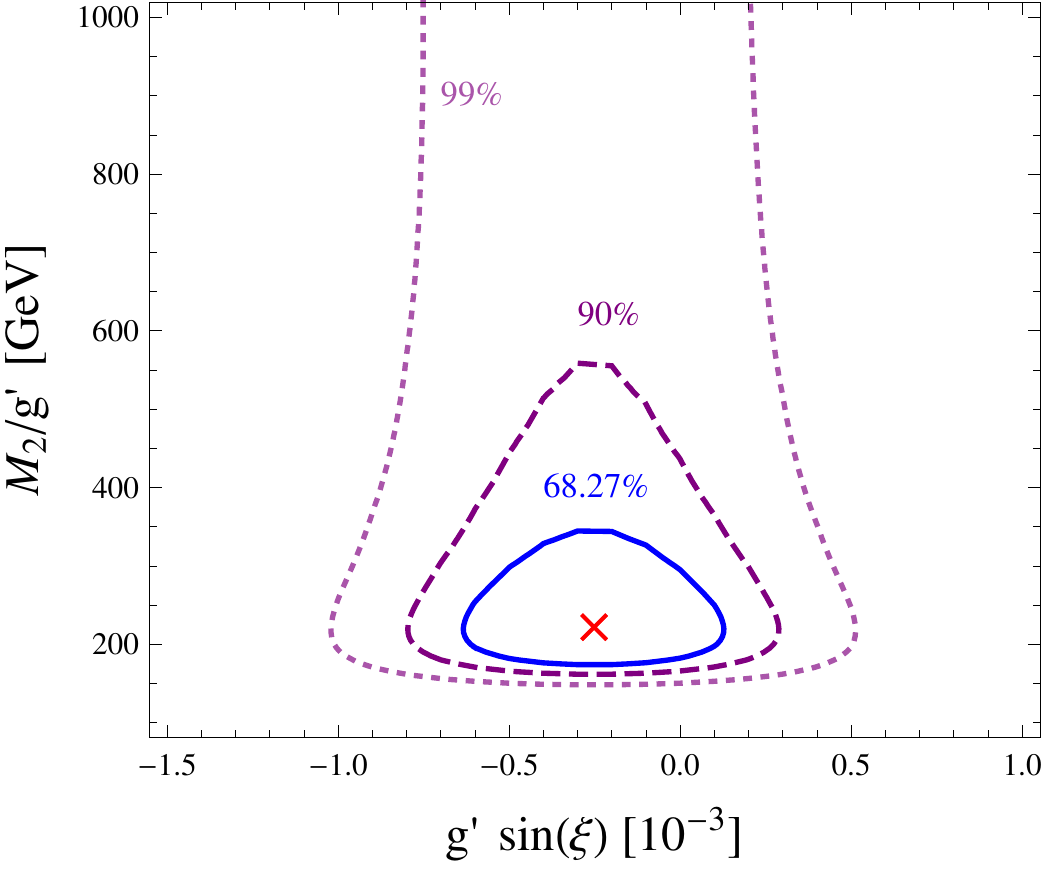}
    \caption{Left: $\chi^2$ as a function of $M_2/g'$ -- marginalized over $\sin \xi$ -- and the $68.27\%$, $90\%$, and $99\%$ C.L. limits for one parameter. Right: $\chi^2$ contours ($68.27\%$, $90\%$, and $99\%$ C.L. for two parameters) in the $M_2$--$\sin\xi$ plane. The cross marks the best-fit values $(g' \sin \xi ,\, M_2/g') = (-2.5\times 10^{-4},\, \unit[219.6]{GeV})$.}
    \label{fig:mzpvsth}
\end{figure}

The best-fit values for the SM parameters hardly change with the
addition of the $Z'$. As can be seen, the reduced $\chi^2_\mathrm{red}
\equiv \chi^2_\mathrm{min}/N_\dof$ decreases from $43.8/44 \simeq
0.995$ to $36.4/42\simeq 0.867$ with the addition of the two
effective parameters $M_2/g'$ and $g' \sin \xi$, a significant
improvement. Marginalizing over $\sin\xi$ we can visualize the
narrowness of the $\chi^2$-minimum (Fig.~\ref{fig:mzpvsth} (left)).

\begin{table}[tb]
\centering
\small
\begin{tabular}{|l|l|l|}
\hline
 & SM & SM+$Z'$\\
\hline
\hline
$M_1$ $[\unit{GeV}]$ & $91.1877$ & $91.1877$\\
$m_t$ $[\unit{GeV}]$ & $164.0$ & $164.0$\\
$m_b$ $[\unit{GeV}]$ & $4.199$ & $4.200$\\
$m_c$ $[\unit{GeV}]$ & $1.270$ & $1.278$\\
$\alpha_s$ & $0.1183$ & $0.1185$\\
$\Delta \alpha_\mathrm{had}^{(3)} (\unit[1.8]{GeV})$ & $5.75\times 10^{-3}$ & $5.72\times 10^{-3}$\\
$m_H$ $[\unit{GeV}]$ & $114.4$ & $114.4$\\
$M_2/g'$ $[\unit{GeV}]$ & $-$ & $219.6$\\
$g'\sin \xi$ & $-$ & $-2.5\times 10^{-4}$\\
\hline
\hline
$\chi^2_\mathrm{min}/N_\dof$ & $43.8/44$ & $36.4/42$\\
\hline
\end{tabular}
\caption{\label{tab:gapp_bestfit} Fit parameters and their best-fit values in an analysis with/without $Z'$. The masses denote pole masses in the $\overline{\mathrm{MS}}$-scheme.}
\normalsize
\end{table}

In Fig.~\ref{fig:mzpvsth} (right) we show the contours $\Delta \chi^2 = 2.30$, $4.61$ and $9.21$, corresponding to $68.27\%$, $90\%$ and $99\%$~C.L.~for 2 parameters. The best-fit value at $g' \sin \xi = -2.5\times 10^{-4}$ and $M_2/g' =  \unit[219.6]{GeV}$ is shown as well.

As can be seen, there is a preferred area for a $Z'$ around $M_2/g' =
200$--$\unit[300]{GeV}$, mainly constrained by $\Delta a_\mu$. 
Performing a separate minimization for each of the parameters (marginalizing over the others), we derive the following $90\%$~C.L.~bounds:
\begin{align}
\begin{split}
	\unit[160]{GeV} \lesssim M_2/g' \lesssim \unit[560]{GeV} \,,\\
	-0.0008 < g' \sin \xi < +0.0003  \,.
\end{split}
	\label{eq:90clbounds}
\end{align}
Going back to the NSI parameters~in Eq.~\eqref{eq:epsilon_mumu_earth}
once more, we can see that $\epsilon^\oplus_{\mu\mu}$ is maximal for
$M_2/g'$ at the lowest bound and $|g'\sin \xi|$ at the largest bound,
resulting in the expression
\begin{align}
	|\epsilon^\oplus_{\mu\mu}| \lesssim 4\times 10^{-4} /g'^2 \,.
\end{align}
Consequently, sizable NSI of order $10^{-2}$ can be generated at the
edge of the allowed $90\%$ C.L.~parameter space, e.g.~for $g'\sim
0.25$--$0.35$, without being in conflict with the direct detection limit.

\subsection{Detection Possibilities at the LHC}
\label{sec:lhc}
The direct detection of the unmixed $Z'$ has already been discussed in
Refs.~\cite{Ma,Baek}, where the most interesting process has been
identified:\footnote{A $3 \mu + \nu$ final state is also of interest,
see Ref.~\cite{Ma}.}

\begin{align}
	p p \ra Z^* X \ra \mu\mu Z' X\,,
	\label{eq:bestprocessever}
\end{align}
i.e.~$Z'$-radiation off final-state muons (or tauons, see
Fig.~\ref{fig:ppto4muon}). This makes the final states $4\mu$, $4\tau$
and $2\mu 2\tau$ especially interesting since the invariant mass
distribution of the lepton-pairs can be checked for a Breit-Wigner
peak of $Z'$. The inclusion of $Z$--$Z'$ mixing does not change these
prospects of direct detection, because the smallness of the mixing
angle $\xi$ as constrained in Eq.~\eqref{eq:90clbounds} reduces the Drell-Yan-production of the
$Z'$ by a factor of $\xi^2$ compared to $Z$ (further suppressed by a
possibly higher $Z'$ mass). Consequently, the rates for $\ell \bar
\ell$ production become nonuniversal at level $\xi^2 \lesssim
10^{-6}$, unlikely to be observed.

\begin{figure}[t]
\centering{
\includegraphics[scale=0.8]{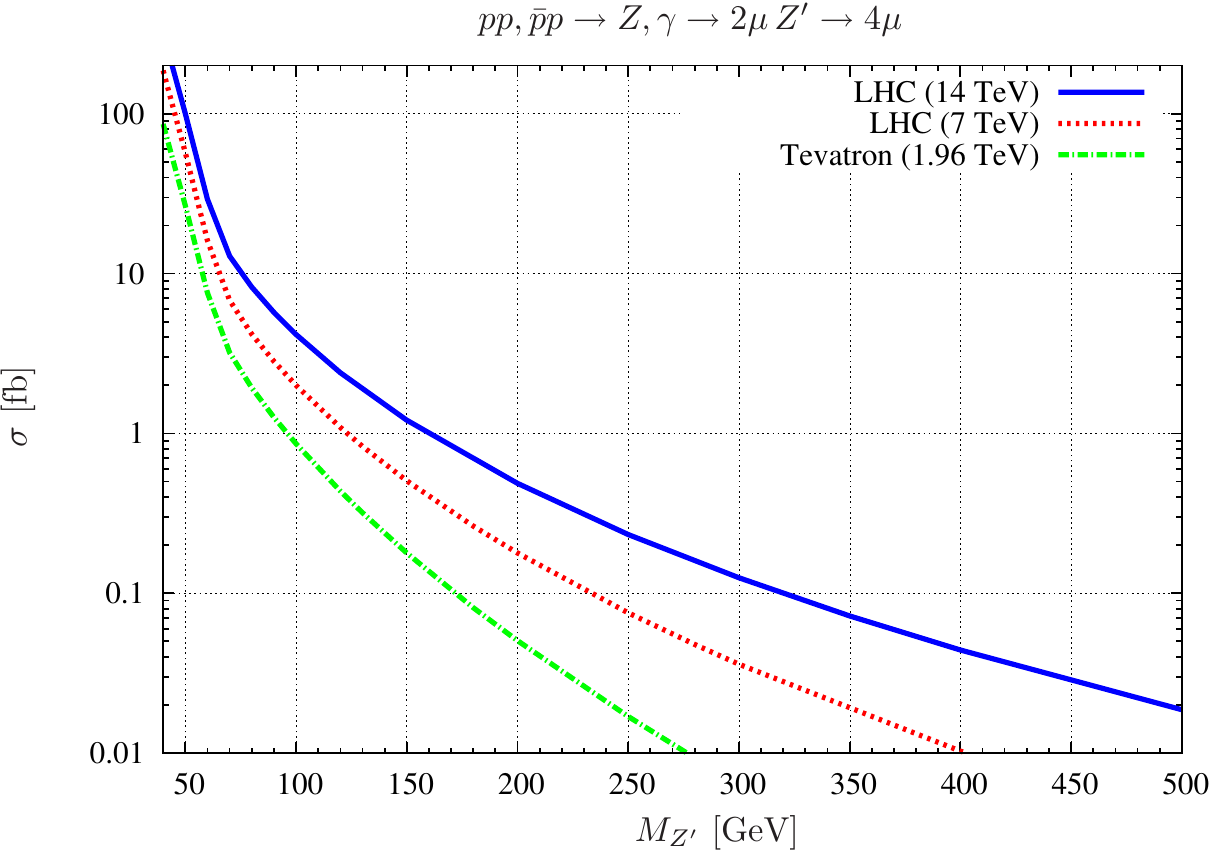}
    \caption{Cross section for the process $p p (p\overline{p}) \ra Z,\gamma \ra 2\mu\, Z' \ra 4 \mu$ with the cuts $p_T >\unit[10]{GeV}$ and $|\eta|<2.5$ for different center-of-mass energies.}
    \label{fig:lhc}}
\end{figure}

Figure~\ref{fig:lhc} shows the cross section for the
process~\eqref{eq:bestprocessever} with four muons in the final state
(using tauons makes no real difference) for the energies
$\sqrt{s}=\unit[7]{TeV}$ and $\unit[14]{TeV}$, as calculated with
CompHEP~\cite{comphep}. The cross section is shown for $g'=1$ and
scales like $g'^2$. The expected integrated luminosity of $\sim
\unit[5]{fb^{-1}}$ until 2012 corresponds to a discovery limit around
$M_{Z'} = \unit[100]{GeV}$ at the LHC ($\sqrt{s}=\unit[7]{TeV}$) for
$g'=1$.\footnote{Here we define discovery by $10$ $Z'$-induced
events.} This is on par with that of Tevatron, due to their higher
luminosity of about $\unit[10]{fb^{-1}}$. Should the LHC be able to
gather $\unit[10]{fb^{-1}}$ until their shutdown in 2013, this limit
can be pushed to about $M_{Z'} = \unit[130]{GeV}$, which is still not
enough to access the interesting parameter space of the global
fit~\eqref{eq:90clbounds}. In the final LHC stage
($\sqrt{s}=\unit[14]{TeV}$, $\L = \unit[100]{fb^{-1}}$) we can probe
the model up to $M_{Z'} = \unit[350]{GeV}$, so this solution to the
anomalous magnetic moment of the muon can be partly tested within the next couple of years. A combined analysis of $4\mu$, $4\tau$ and $2\mu 2\tau$ final states can be used to increase statistics and improve these discovery limits.

The discovery potential for a linear $e^+ e^-$-collider with center-of-mass energy $\sqrt{s} = \unit[0.5]{TeV}$ ($\sqrt{s} = \unit[1]{TeV}$) has been calculated in Ref.~\cite{Ma} to be $M_{Z'} = \unit[300]{GeV}$ ($M_{Z'} = \unit[500]{GeV}$) for the coupling constant $g'=1$. A muon collider would, of course, be the ideal experiment to test this model, since precision measurements could tighten the bounds on $Z'$ even for $\sqrt{s}< M_{Z'}$.

\section{Simple Model for the Neutrino Sector}\label{sec:neutrino_sector}

We extend the fermion content of our theory, without introducing anomalies, by three right-handed neutrinos in the representations of $G_\sm \times \Uprime$
\begin{align}
	N_1 \sim (\vec{1},\vec{1},0)(0) \,, &&
	N_2 \sim (\vec{1},\vec{1},0)(+1) \,, &&
	N_3 \sim (\vec{1},\vec{1},0)(-1) \,.
\end{align}
The gauge-invariant Yukawa couplings are (with $N_i^T \mathcal{C}^{-1}N_j = - N_i^T \mathcal{C} N_j = - \overline{N}_i^c N_j$)
\begin{align}
	-\L \supset \, -\frac{1}{2} N_i^T \mathcal{C}^{-1} (\mathcal{M}_{R})_{ij} N_j 
	+ \lambda^{e 1}_H \overline{L}_e \tilde{H} N_1
	+\lambda^{\mu 2}_H \overline{L}_\mu \tilde{H} N_2
	+\lambda^{\tau 3}_H \overline{L}_\tau \tilde{H} N_3
	+\hc  \,,
\end{align}
where $\mathcal{M}_R$ has the $L_\mu-L_\tau$-symmetric structure
\begin{align}
	\mathcal{M}_R = \matrixx{X & 0 & 0\\ 0 & 0 & Y\\ 0 & Y & 0} .
	\label{eq:mutaumajoranamatrix}
\end{align}
Electroweak symmetry breaking ($H \rightarrow (0,\, \frac{1}{\sqrt{2}} (h + v))^T$) generates the bilinear terms
\begin{align}
	\L \supset - \matrixx{\overline{\nu}_e & \overline{\nu}_\mu & \overline{\nu}_\tau} \matrixx{m_{\nu_e} & 0 & 0\\ 0 & m_{\nu_\mu} & 0 \\ 0 & 0 & m_{\nu_\tau}} \matrixx{N_1\\ N_2\\ N_3}
	+\hc  \,,
\label{eq:neutrino_majorana_massterm}
\end{align}
where we introduced the Dirac mass matrix $m_D$ with the entries $m_{\nu_i} \equiv -\lambda^{i}_H v/\sqrt{2}$.
Invoking the seesaw mechanism \cite{seesaw} in the form of 
$X,Y \gg m_{\nu_i}$ results in the low-energy mass matrix
\begin{align}
	\mathcal{M}_\nu \simeq - m_D \mathcal{M}_R^{-1} m_D^T 
	= -\matrixx{\frac{m_{\nu_e}^2}{X} & 0 & 0\\ \cdot & 0 & \frac{m_{\nu_\mu} m_{\nu_\tau}}{Y}\\ \cdot & \cdot & 0}  ,
\label{eq:symmetric_majorana_matrix}
\end{align}
while the mass matrix of the charged leptons is diagonal because electron, muon and tauon carry different $U(1)'$ charges.
We stress here that $X$, $Y$ as well as $m_{\nu_\mu}$, $m_{\nu_\tau}$
and $m_{\nu_e}$ are allowed by the symmetry, and hence expected to be
of similar magnitude each. 
The eigenvalues of Eq.~(\ref{eq:symmetric_majorana_matrix}) are
$-m_{\nu_e}^2 / X$ and $\pm m_{\nu_\mu} m_{\nu_\tau} / Y$ and are
therefore naturally of similar magnitude, i.e.~there is at most a mild
hierarchy between the neutrino masses. This is what observations seem
to tell us, as the neutrino mass hierarchy is much weaker than the one
of charged leptons or quarks. 
 The atmospheric mixing angle $\theta_{23}$ associated with this mass
matrix is maximal, while the other two mixing angles $\theta_{13}$ and
$\theta_{12}$ are zero, and hence will only be induced by breaking the
$\Uprime$ symmetry. The degenerate neutrino pair $\pm m_{\nu_\mu}
m_{\nu_\tau} / Y$ will also be split by the breaking. 

The phenomenology of texture zeros in neutrino mass matrices (like in
Eq.~\eqref{eq:symmetric_majorana_matrix}) has been discussed, for
example, in Ref.~\cite{texturezeros}, where a classification for the
different structures is given. Most importantly, an analysis shows
that $\mathcal{M}_\nu$ can have at most two texture zeros to be
phenomenologically successful. 	
There is, of course, no unique way to break the $\Uprime$ symmetry
spontaneously. Even the restriction to a seesaw-I implementation
allows for various models with very different phenomenology. The
choice to break the symmetry in the neutrino mass matrix either in the
Dirac matrix $m_D$ or in the right-handed matrix $\mathcal{M}_R$ fixes
at least the $G_\sm$ quantum numbers of the scalar fields to $SU(2)_L$
doublets or singlets, respectively. Let us focus on doublets for a
moment: To resolve the magnetic moment of the muon $\Delta a_\mu$, we
need $M_{Z'}/g' = \sqrt{\sum_j {Y'_j}^2 v_j^2} \gtrsim
\unit[200]{GeV}$, where $Y'_j$ denotes the $\Uprime$ charge of the
scalar $\phi_j$ with VEV~$v_j$. Introducing just one additional
doublet would result in a large $Z$--$Z'$ mixing angle, so we have to
introduce another doublet with opposite $Y'$ (this model was used in
Ref.~\cite{Ma}). With mild fine-tuning (or an additional symmetry) we
can ensure the smallness of the off-diagonal $Z$--$Z'$-mass element
$\delta \hat M^2 \sim g_Z \, g' (v_1^2 - v_2^2)$. However, since all the
doublets contribute to $M_Z$ and $M_{W^\pm}$, we have the additional
constraint $\sum_\mathrm{doublets} v_j^2 \simeq
(\unit[246]{GeV})^2$. This leaves at most $\unit[140]{GeV}$ for the
VEV of the Standard Model Higgs doublet, which is however the only
doublet that couples to the top-quark. From these remarks it is clear
that this three-doublet model cannot describe the whole interesting
parameter space~of Eq.~\eqref{eq:90clbounds}.

Breaking $\Uprime$ solely in the right-handed sector via SM-singlets
is very simple and allows for an arbitrary large $Z'$ mass, but has
little interesting phenomenology, because these scalars dominantly
couple to the heavy neutrinos and $Z'$, both already difficult to
probe. We therefore opt for a combined breaking, using doublets and
singlets, as this model displays numerous interesting effects that can
be tested experimentally.
To that effect, we introduce another scalar doublet $\phi = (\phi^+,\,
\phi^0)^T \sim (\vec{1},\vec{2},+1)(-1)$ and one SM-singlet $S \sim
(\vec{1},\vec{1},0)(-1)$, leading to the following additional neutrino interactions
\begin{align}
\begin{split}
 	-\L\  \supset\  &+ \lambda^{e 3}_\phi \overline{L}_e \tilde \phi N_3 + \lambda^{\mu 1}_\phi \overline{L}_\mu \tilde \phi N_1 +\lambda^{12}_S \overline{N}_1^c N_2 S +\lambda^{13}_S \overline{N}_1^c N_3 \overline S +\hc
\end{split}
\end{align}

 If $\phi^0$ acquires a VEV it generates entries in $m_D$ ($d\equiv
-\lambda_\phi^{e3} \langle \phi^0 \rangle$, $f\equiv
-\lambda_\phi^{\mu 1} \langle \phi^0 \rangle$), while a nonzero
$\langle S \rangle$ modifies $\mathcal{M}_R$ via entries $s\equiv
\lambda_S^{1 2} \langle S\rangle$ and $t\equiv \lambda_S^{1 3} \langle S\rangle$:
\begin{align}
	m_D = \matrixx{m_{\nu_e} & 0 & d\\ f & m_{\nu_\mu} & 0\\ 0 & 0 & m_{\nu_\tau}} , &&
	\mathcal{M}_R = \matrixx{X & s & t\\ s & 0 & Y \\ t & Y & 0} ,
\end{align}
so the low-energy neutrino mass matrix in linear order takes the form
\begin{align}
 \mathcal{M}_\nu \simeq  \matrixx{
 -\frac{m_{\nu_e}^2}{X}  & 
 -\frac{m_{\nu_e} f}{X}-\frac{m_{\nu_\mu} d}{Y} +\frac{m_{\nu_e}m_{\nu_\mu} t }{X Y} & 
 \frac{m_{\nu_e}m_{\nu_\tau}s}{XY}\\
 \cdot & 
 0 & 
 -\frac{m_{\nu_\mu} m_{\nu_\tau}}{Y} \\
 \cdot & 
 \cdot & 
 0} ,
\end{align}
where we used $s,t \ll X,Y$ and $d,f\ll m_{\nu_\alpha}$. It is important to note that the small parameters $s$ and $t$ do not spoil the validity of the seesaw mechanism, since the light masses are still always suppressed by $X$ or $Y$. The remaining texture zeros break $\Uprime$ by two units and are therefore filled by terms of order two in our perturbative expansion.
We can reduce the number of parameters by one with the introduction of
\begin{align}
	a = \frac{m_{\nu_\tau}}{m_{\nu_e}} \,, \qquad
	b = \frac{m_{\nu_\mu}}{m_{\nu_e}} \,, \qquad
	c = 1-k =  \frac{X}{Y} \,,\\
	\epsilon = \frac{s}{Y}  \,,\qquad
	\alpha = \frac{t}{Y}  \,,\qquad
	\delta = \frac{d}{m_{\nu_e}} \,, \qquad
	\gamma = \frac{f}{m_{\nu_e}}\,.\label{eq:small_parameters}
\end{align}
Small breaking of $\Uprime$ and quasidegenerate masses correspond to $\epsilon$, $\alpha$, $\delta$, $\gamma$, $k\ll 1$ and $a\, b \simeq 1$, so we decompose
\begin{align}
	\mathcal{M}_\nu / \left(\frac{m_{\nu_e}^2}{Y}\right) =
\mathcal{M}_\nu^0 + \Delta \mathcal{M} = 
- \matrixx{1 & 0 & 0\\ 0 & 0 & a\, b\\ 0 & a\, b & 0} + \Delta \mathcal{M}\,,
\end{align}
with the symmetric perturbation matrix (without any approximations)
\begin{align}
	\Delta \mathcal{M} = \matrixx{-\delta\, \epsilon\, (-2 + \delta\, \epsilon) & b\, \alpha \,(1+\delta\, \epsilon) + \gamma\, (-1 + \delta\, \epsilon) + b\, \delta (k-1) & a\,\epsilon (1-\delta\, \epsilon)\\
	\cdot & - (\gamma - b\, \alpha)^2 & a\, (b \,\alpha \,\epsilon + \epsilon\,\gamma +b \,k)\\
	\cdot & \cdot & -a^2 \epsilon^2} ,
\end{align}
which shifts the eigenvalues of $\mathcal{M}_\nu^0$ from $-1$ and $\pm a\, b (1-k)$ to 
\begin{align}
\begin{split}
	\lambda_1 &\simeq -1 + 2 \delta\, \epsilon \,, \\
	\lambda_{2,3} &\simeq \pm a\, b (1-k) \mp a\, \epsilon\, \gamma \mp a\, b\, \alpha \,\epsilon - \frac{1}{2} b^2 \alpha^2- \frac{1}{2} a^2 \epsilon^2 + b \,\alpha\, \gamma  - \frac{1}{2} \gamma^2\,.
\end{split}
\end{align}
Assuming a scale ${m_{\nu_e}^2}/{Y}\sim \unit[0.1]{eV}$, the
atmospheric mass-squared difference $\Delta m_\mathrm{atm}^2 \simeq
\unit[2.4\times 10^{-3}]{eV^2}$ will be generated by $|a\,b| = 1 +
\mathcal{O}(0.1)$, while the solar one goes quadratic in the small
parameters in Eq.~\eqref{eq:small_parameters} and hence one needs them to be
of order $\mathcal{O}(0.05)$. The mixing angle $\theta_{13}$ will be
small but nonzero, since it is linear in the small parameters and can
be further increased by proper values of $a$ and $b$: 
\begin{align}
	\sin \theta_{13} \simeq \frac{1}{\sqrt{2} (1+a\,b)} \, \left( a \,\epsilon + \gamma + b\,(\delta -\alpha)\right) \sim 10^{-1}\textrm{--}10^{-3}\,,
\end{align}
in agreement with recent T2K results~\cite{t2k} and global fits~\cite{th13fit}.
The deviation from $\sin \theta_{23}=1/\sqrt{2}$ on the other hand is quadratic in the small parameters and typically of order $\mathcal{O}(10^{-2})$:
\begin{align}
	\Delta \sin \theta_{23} \simeq \frac{1}{4 \sqrt{2}} \left( a^2 \epsilon^2 - b^2 \alpha^2 + 2\, b\, \alpha\, \gamma -\gamma^2\right)\,.
\end{align}
These deviations from $\mu$-$\tau$ symmetry can be checked by future
experiments, but are, of course, not specific to this model. As a
consequence of the partial or quasidegeneracy the absolute neutrino masses are
rather large, which, due to the Majorana nature of the light
neutrinos, allows for neutrinoless double $\beta$-decay in the reach
of upcoming experiments. Since there are no particularly
model-specific predictions, we omit a further discussion of
$0\nu\beta\beta$ and direct mass-measurement experiments.

From Sec.~\ref{sec:gappfit} we already know the favored values for the $\Uprime$ breaking scale $s,t\sim \langle S\rangle \sim M_{Z'}/g' \sim \unit[200]{GeV}$, which puts the $N_R$ scale in the range $X,Y \simeq s/\epsilon \sim 1$--$\unit[10]{TeV}$.

\section{Details of the Scalar Sector}
\label{sec:scalar_sector}

As already mentioned in Section \ref{sec:neutrino_sector}, we introduce the scalar fields 
\begin{align}
 H = \matrixx{h^+ \\ h^0} \sim (\vec{1},\vec{2},+1)(0) \,,&&
\phi = \matrixx{\phi^+\\ \phi^0} \sim (\vec{1},\vec{2},+1)(-1) \,,&&
S \sim (\vec{1},\vec{1},0)(-1)\,,
\end{align}
which adds up to $10$ real scalar fields. Four of these will serve as
the longitudinal modes of the $W^\pm$, $Z$ and $Z'$ bosons, so we will
end up with $6$ physical scalar fields (instead of one in the
SM). Since we introduce an additional Higgs doublet, the phenomenology
of the scalars will be similar to the usual two-Higgs-doublet models
(2HDM, see Ref.~\cite{Branco:2011iw} for a recent review). The general
potential for our fields can be written as
\begin{align}
\begin{split}
 	V = &-\mu_1 |H|^2 +\lambda_1 |H|^4 - \mu_2 |\phi|^2 + \lambda_2 |\phi|^4 -\mu_3 |S|^2 + \lambda_3 |S|^4\\
&+\delta_1 |H|^2 |\phi|^2 + \delta_2 |H^\dagger \phi|^2 + \delta_3 |H|^2 |S|^2 + \delta_4 |\phi|^2 |S|^2\\
&-\left( \sqrt{2}|\mu| e^{i\kappa} H^\dagger \phi \overline{S} +\hc\right)\,.
\end{split}
\label{eq:scalar_potential_1s1d}
\end{align}
The positivity of the potential gives constraints on the coefficients, since we have to ensure that there is a minimum around which we can use perturbation theory. To this effect, one studies the different directions in field-space (e.g.~$S= 0$ and $|\phi|,|H| \ra \infty$) to find a number of algebraic equations that ensure $V>0$.  In the case of Eq.~\eqref{eq:scalar_potential_1s1d}, the quartic part of the potential (the relevant part for large field values, i.e.~the limit $H, \phi, S\ra \infty$) has the structure of a quadratic form as long as $\delta_2=0$ or the field direction satisfies $H^\dagger \phi = 0$, which then allows us to simply use Sylvester's criterion for positive-definite quadratic forms to determine the relevant conditions: 
\begin{align}
\begin{split}
 	0&< \lambda_j \,, \qquad \qquad 0< 4 \lambda_1 \lambda_2 - \delta_1^2 \,,\\
0&< 4\lambda_1 \lambda_2 \lambda_3 +\delta_1 \delta_3 \delta_4 - \lambda_1 \delta_4^2 - \lambda_2 \delta_3^2 - \lambda_3 \delta_1^2 \,.
\end{split}
\end{align}
Including $\delta_2$ yields the supplementary bound
\begin{align}
\begin{split}
0&< 4\lambda_1 \lambda_2 \lambda_3 +(\delta_1+\delta_2) \delta_3 \delta_4 - \lambda_1 \delta_4^2 - \lambda_2 \delta_3^2 - \lambda_3 (\delta_1+\delta_2)^2 \,.
\end{split}
\end{align}
Additional constraints come from the positivity of the scalar masses, which are however more intricate and will not be explicitly derived here; neither will the bounds from perturbativity and unitarity, which, in principle, give upper bounds on the couplings.
Introducing the VEVs\footnote{While we choose all VEVs real and positive for simplicity, it must be stressed that this is not the most general case.} 
\begin{align}
 \re S \ra \langle S \rangle \equiv v_S/\sqrt{2}\,, && \re \phi^0 \ra \langle \phi^0 \rangle \equiv v_{\phi}/\sqrt{2}\,, && \re h^0 \ra v/\sqrt{2}\,,
\end{align}
and minimizing the potential, gives three equations for $\mu_{1,2,3}$, which we plug back into the potential. To calculate the masses we will go to unitary gauge, i.e.~eliminate the unphysical degrees of freedom, as determined by the kinetic terms:
\begin{align}
\begin{split}
 	\L &\supset (D_\mu H)^\dagger (D^\mu H) + (D_\mu \phi)^\dagger (D^\mu \phi) +(D_\mu S)^\dagger (D^\mu S)\\
&= | \del_\mu h^0  -i\frac{e}{2s_W c_W} Z_\mu h^0 -i\frac{e}{\sqrt{2} s_W} W^-_\mu h^+|^2\\
&\quad +|\del_\mu h^+ - i\frac{e}{s_W c_W}(c_W^2 -s_W^2) Z_\mu h^+ - i e A_\mu h^+ -i\frac{e}{\sqrt{2}s_W} W^+_\mu h^0 |^2\\
&\quad  +| \del_\mu \phi^0 + i g' Z'_\mu \phi^0 -i\frac{e}{2s_W c_W} Z_\mu \phi^0 -i\frac{e}{\sqrt{2} s_W} W^-_\mu \phi^+|^2\\
&\quad +|\del_\mu \phi^+ +i g' Z'_\mu \phi^+ - i\frac{e}{s_W c_W}(c_W^2 -s_W^2) Z_\mu \phi^+ - i e A_\mu \phi^+ -i\frac{e}{\sqrt{2}s_W} W^+_\mu \phi^0 |^2\\
&\quad  +| \del_\mu S + i g' Z'_\mu S |^2\,.
\end{split}
\label{eq:gauge_interactions_1s1d}
\end{align}
Expanding the fields around the VEVs, we find the mass terms for the gauge bosons 
\begin{align}
\begin{aligned}
 	M_W^2 &= \frac{e^2}{4 s_W^2} (v^2+v_{\phi}^2)\,, &\qquad
	M_Z^2 &= \frac{e^2}{4 s_W^2 c_W^2} (v^2+v_{\phi}^2)\,, \\
	M_{Z'}^2 &= g'^2 (v_{\phi}^2 +v_S^2)\,, &\qquad
	\delta \hat M^2 &= -\frac{e}{2 s_W c_W} g' v_{\phi}^2 \,.
\end{aligned}
\end{align}
Small $Z$--$Z'$ mixing demands a small VEV $v_\phi$, but since the
mixing angle $\xi $ in Eq.~(\ref{eq:abc}) is quadratic in the VEVs, this only constrains $v_\phi \lesssim \unit[10]{GeV}$ (using the limits~\eqref{eq:90clbounds} and assuming $M_{Z'}>M_Z$, i.e.~$g'\sim 1$). The main contribution to the $Z'$ mass has to come from $v_S$, the anomalous magnetic moment $\Delta a_\mu$ gives the constraint $\sqrt{v_S^2 + v_{\phi}^2} \gtrsim \unit[200]{GeV}$. In the following, approximations are made with the scaling $v_\phi \ll v_S \sim v$.

Aside from the mass terms, we also find the cross terms between gauge bosons and Goldstone bosons, namely: 
\begin{align}
\begin{split}
 	\L \supset &- \frac{e}{\sqrt{2} s_W c_W} Z_\mu \del^\mu (v_{\phi}\, \im \phi^0 + v\, \im h^0) +\sqrt{2} g' Z'_\mu \del^\mu (v_{\phi}\, \im \phi^0 + v_S\, \im S) \\
&-\frac{e}{2 s_W} i W^+_\mu \del^\mu (v_{\phi}\, \phi^- + v\, h^-) +\hc 
\end{split}
\end{align}
We read off the Goldstone fields (not properly normalized)
\begin{align}
 	G^- \sim v_{\phi}\,  \phi^- + v\,  h^-, && G_Z \sim v_{\phi}\,  \im \phi^0 + v\,  \im h^0, && G' \sim v_{\phi}\, \im \phi^0 +v_S\,  \im S\,,
\end{align}
which are not orthogonal. Using the gauge freedom to fix $G^- = G_Z =
G' = 0$ would result in physical scalars with unconventional kinetic
terms; instead of rotating the noncanonical kinetic terms, 
this can be avoided also by first constructing a orthonormal basis
from $G_Z$ and $G'$. 
We define the physical field $\sigma$ via $\sigma \sim G' \times G_Z$,\footnote{To make use of the cross product we identify $G'$ and $G_Z$ with vectors in the basis $(\im S, \im \phi^0, \im h^0)$.}
then ``rotate" $G_Z$ to $\tilde G_Z \equiv \sigma \times G'$. These
fields are connected to the gauge eigenstates by a unitary
transformation:
\begin{align}
\begin{split}
	\matrixx{G'\\\tilde G_Z \\ \sigma} &= \matrixx{\cos \theta & \sin\theta &0\\ -\cos\beta \sin\theta & \cos\beta \cos\theta & \sin\beta\\ \sin\beta \sin\theta & -\sin\beta \cos\theta & \cos\beta} \matrixx{\im S\\\im \phi^0 \\\im h^0}
\end{split}
\end{align}
with the two angles
\begin{align}
 	\tan \theta \equiv \frac{v_{\phi}}{v_S}\,, &&
	\tan \beta \equiv \frac{v}{v_S v_{\phi}}\sqrt{v_S^2+v_{\phi}^2} = \frac{v}{v_S \sin\theta}\,.
	\label{eq:thetaandbeta}
\end{align}
The expected scaling $v_{\phi} \ll v_S \sim v$ implies $\sin\theta, \cos\beta \ll 1$. The unitary gauge, $G' = \tilde G_Z =0$, leaves the physical field $\sigma$, contributing to the potential through
\begin{align}
 	\matrixx{\im S\\\im \phi^0 \\\im h^0} = \matrixx{\cos \theta & \sin\theta &0\\ -\cos\beta \sin\theta & \cos\beta \cos\theta & \sin\beta\\ \sin\beta \sin\theta & -\sin\beta \cos\theta & \cos\beta}^T \matrixx{0\\ 0 \\ \sigma} 
= \matrixx{\sin\beta\sin\theta \,\sigma\\ -\sin\beta \cos\theta \,\sigma\\ \cos\beta \,\sigma} .
	\label{eq:neutral_goldstones}
\end{align}
The field $\sigma$ consists mainly of $\im\phi$, so the imaginary part of $h^0$ is not zero as in the SM, but suppressed by $\cos\beta$.
The charged Goldstone boson is easier to handle, we have
\begin{align}
 	\matrixx{\phi^-\\ h^-} = \matrixx{\cos \beta^- & -\sin\beta^- \\ \sin\beta^- & \cos\beta^-}\matrixx{G^-\\\sigma^-} \stackrel{G^-\ra \ 0}{\xrightarrow{\hspace*{1.3cm}}} \matrixx{-\sin\beta^-\, \sigma^-\\ \cos\beta^-\, \sigma^-} ,
 	\label{eq:charged_goldstones}
\end{align}
with the angle $\tan\beta^- \equiv v/v_{\phi} = \cos\theta \tan\beta \simeq \tan\beta \gg 1$.
The physical fields $\sigma$, $\sigma^\pm$, $\re S \equiv S$, $\re \phi^0 \equiv \phi$ and $\re h^0 \equiv h$ are not mass-eigenstates. Setting, for simplicity, the CP-violating angle $\kappa$ in the potential~\eqref{eq:scalar_potential_1s1d} to zero, we can at least read off the masses for $\sigma$ and $\sigma^\pm$:
\begin{align}
 	m_\sigma^2 &= \frac{|\mu| \, v \,  v_S}{v_{\phi}} + \frac{|\mu| \,  v_{\phi} \,  v_S}{v} +\frac{|\mu|  \, v  \, v_{\phi}}{v_S}\,,\\
m_{\sigma^{\pm}}^2 &= \frac{|\mu| \,  v  \, v_S}{v_{\phi}} + \frac{|\mu|  \, v_{\phi} \,  v_S}{v} -\frac{1}{2}\delta_2 (v_{\phi}^2+v^2)\,.
\end{align}
In the approximation we will use extensively, $v_{\phi} \ll v_S \sim v$, only the first term contributes and $m_\sigma \simeq m_{\sigma^\pm}$. The masses $m_\sigma$ and $m_{\sigma^{\pm}}$ increase for decreasing $v_{\phi}$, reminiscent of an inverse seesaw mechanism; useful values for $|\mu|$ will be discussed below.
The CP-even scalars share the symmetric mass matrix (in $(S,\phi,h)$ basis)
\begin{align}
 	\mathcal{M}_{\mathrm{CP-even}}^2 = \matrixx{2\lambda_3 v_S^2 + \frac{|\mu| \,  v \,  v_{\phi}}{s} & -|\mu| v + \delta_4 v_{\phi} v_S & -|\mu| v_{\phi}+\delta_3  \, v v_S\\
\cdot & 2 \lambda_2 v_{\phi}^2 + \frac{|\mu|  \, v  \, v_S}{d} & -|\mu| v_S+ (\delta_1 + \delta_2) v v_{\phi}\\
\cdot & \cdot & 2 \lambda_1 v^2 + \frac{|\mu| \,  v_{\phi}  \, v_S}{ v}} ,
\end{align}
with the approximate eigenvalues (labeled according to the predominant field in the unmixed scenario)
\begin{align}
 	m_{\tilde h,\tilde S}^2 &\simeq \lambda_1 v^2 + \lambda_3 v_S^2 \pm \sqrt{(\lambda_1 v^2 - \lambda_3 v_S^2)^2 + \delta_3^2 \,  v_S^2  \, v^2} \,,\\
 m_{\tilde \phi}^2 &\simeq \frac{|\mu|  \, v  \, v_S}{v_{\phi}} \,.
\end{align}
For small $\cos\beta\simeq v_{\phi}/v$, the fields $\sigma$, $\sigma^+$ and $\phi$ have degenerate masses. In contrast to other 2HDM, we do not have a light pseudoscalar $\sigma$, because it has roughly the same mass as the charged scalar $\sigma^\pm$. In the next section, we will see that the charged scalar mass is bounded from below, $m_{\sigma^\pm} \gtrsim \unit[80]{GeV}$, so there are no decay modes of $h$ into real $\sigma\sigma$, $\sigma^+ \sigma^-$ or $\phi\phi$, unless $h$ has a mass of at least $\unit[160]{GeV}$. $S$~could, in principle, have a mass low enough to allow $h\ra SS$, depending on $\lambda_3$. Decay channels as signatures in collider experiments will be discussed below. We point out that without large mass splittings in the scalar sector, the quantum corrections to the $\rho$ parameter, due to scalar loops, will be small~\cite{Grimus:2007if}. 

Obviously, the $\mu$ term in the potential is crucial for the
generation of large scalar masses, without it, we would end up with
masses $\sim v_{\phi}$, either below $M_Z$ (introducing new invisible
decay channels for $Z$) or above it (introducing too large of a
$Z$--$Z'$-mixing angle). The potential in the $(\sigma, \sigma^\pm,
S,\phi,h)$ basis is ridiculously lengthy and will not be shown
here. It involves the interaction terms given in
Table \ref{tab:1s1d_interactions}; also shown are the interactions with
the gauge bosons. Making $|\mu|$ and the $\delta_i$ small,
e.g.~$|\mu|\sim v_{\phi}$, results in small mass mixing of order
$\mu/v$ and $\delta_3$; for simplicity, we will work in zeroth order
and treat $S$, $\phi$ and $h$ as mass eigenstates. The additional mass
mixing can be of the same order as the mixing through $\beta$ and
$\beta^-$, consequently the combined mixing could be larger or
smaller, depending on their relative sign in the coupling, similar to
usual 2HDM~\cite{Branco:2011iw}. Since we are only performing order of
magnitude approximations in the scalar sector, 
we do not go into more detail.

Just as an aside, we mention that none of the scalars are stable. The
scalars $\sigma$ and $\sigma^-$ couple directly to fermions (albeit
weakly) and will decay through such channels. The scalars $\phi$ and
$S$ couple predominantly to the heavy neutrino sector, but can in any
way decay via a $Z' Z'$. So, without invoking some additional discrete
symmetries, this model provides no candidate for dark matter.

\begin{table}[t]
\centering
	\renewcommand{\baselinestretch}{1.15}\normalsize 
\small
\begin{tabular}{|l|l|l|l|l|l|l|}
\hline
$h^3$ & 
$h^4$ & 
$S^3$ &
$S^4$ & 
$\phi^3$ & 
$\phi^4$ &
$\sigma^4$ \\ 
$\sigma^2 h$ & 
$h\phi S$ &
$\sigma^2 \phi$ & 
$\sigma^2 S$ &
$\phi^2 S^2$ & 
$h^2 S^2$ & 
$h^2 \phi^2$ \\
$\sigma^2 \phi^2$ & 
$\sigma^2 h^2$ & 
$h^2 S$ &
$h^2 \phi$ & 
$\phi^2 h$ & 
$\sigma^2 S^2$ &
$\phi^2 S$ \\ 
$S^2 \phi$ &
$h S^2$ & 
$\sigma^+ \sigma^- \sigma^+ \sigma^-$ & 
$h^2 \sigma^+ \sigma^-$ & 
$\phi^2 \sigma^+ \sigma^-$ &
$\sigma^2 \sigma^+ \sigma^-$ & 
$h \sigma^+ \sigma^-$ \\
$\phi \sigma^+ \sigma^-$ & 
$S \sigma^+ \sigma^-$ & 
$h \phi \ \sigma^+ \sigma^-$ & 
$S^2 \sigma^+ \sigma^-$ &
 & & \\
\hline
\hline
$Z^2 h$ &
$W^+ W^- \phi$ & 
$Z^2 \sigma^+ \sigma^-$ & 
$Z A \sigma^+ \sigma^-$ & 
$Z W^- \sigma^+ \phi$ &
$A^2 \sigma^+ \sigma^-$ & 
$Z^2 h^2$ \\ 
$W^- A \sigma^+ \phi$ & 
$W^+ W^- \sigma^2$ & 
$W^+ W^- \phi^2$ & 
$Z W^- \sigma^+ \sigma$ & 
$W^+ A \sigma^- \sigma$ & 
$W^+ W^- \sigma^+ \sigma^-$ &
$Z W^- \sigma^+ h$ \\ 
$W^- A \sigma^+ h$ & 
$W^+ W^- h$ & 
$W^+ W^- h^2$ &
$Z^2 \sigma^2$ & 
$Z^2 \phi$ &
$Z^2 \phi^2$ &
 \\
\hline
\hline
$Z'^2 \phi^2$ & 
$Z'^2 S^2$ & 
$W^- Z' \sigma^+ \phi$ & 
$Z Z' \sigma^+\sigma^-$ & 
$A Z' \sigma^+ \sigma^-$ &
$W^+ Z'\sigma^- \sigma$ & 
$Z'^2 \sigma^+\sigma^-$ \\
$W^+ Z' \sigma^-$ & 
$Z'^2 \phi$ & 
$Z'^2 S$ &
$Z'^2 \sigma^2$ & 
$Z Z' \sigma^2$ & 
$Z Z' \phi$ & 
$Z Z'\phi^2$ \\
\hline
\hline
$A\sigma^+ \sigma^-$ &
$Z \sigma^+ \sigma^-$ &
$Z \sigma h$ &
$Z \sigma \phi$ &
$W^- \sigma^+ \sigma$ &
$W^- \sigma^+ h$ &
$W^- \sigma^+ \phi$ \\
$Z' \phi \sigma$ &
$Z' \sigma S$ &
$Z' \sigma^+ \sigma^-$ & & & & \\
\hline
\end{tabular}
\caption{\label{tab:1s1d_interactions} Interaction vertices involving scalars among themselves (first row), with SM vector bosons (second row), with the $Z'$ boson (third row), and couplings to vector bosons involving derivatives (last row).}
\end{table}

\subsection{Yukawa Interactions and Lepton Flavor Violation}
\label{sec:LFV}
The gauge-invariant Yukawa interactions of the two doublets $H$ and $\phi$ and the singlet $S$ to the leptons and right-handed neutrinos $N_i$ can be written as:
\begin{align}
\begin{split}
 	-\L\  \supset\  &+\lambda^{e e}_H \overline{L}_e H e_R + \lambda^{\mu\mu}_H \overline{L}_\mu H \mu_R + \lambda^{\tau\tau}_H \overline{L}_\tau H \tau_R \\
&+\lambda^{e 1}_H \overline{L}_e \tilde H N_1 + \lambda^{\mu 2}_H \overline{L}_\mu \tilde H N_2 + \lambda^{\tau 3}_H \overline{L}_\tau \tilde H N_3\\
&+\lambda^{e\mu}_\phi \overline{L}_e \phi \mu_R + \lambda^{\tau e}_\phi \overline{L}_\tau \phi e_R\\
&+ \lambda^{e 3}_\phi \overline{L}_e \tilde \phi N_3 + \lambda^{\mu 1}_\phi \overline{L}_\mu \tilde \phi N_1 +\lambda^{12}_S \overline{N}_1^c N_2 S +\lambda^{13}_S \overline{N}_1^c N_3 \overline S +\hc
\end{split}
\end{align}
In unitary gauge we replace the scalars by the physical degrees of
freedom $\sigma$~\eqref{eq:neutral_goldstones},
$\sigma^\pm$~\eqref{eq:charged_goldstones}, $S$, $\phi$ and
$h$. Denoting $\sin\beta$ with $s_\beta$,  $\sin\beta^-$ with
$s_\beta^-$ etc., this becomes:
\begin{align}
\begin{split}
 	-\L \ \supset \ 
&\sum_{\ell=e,\mu,\tau} \lambda^{\ell\ell}_H \left[ \frac{v}{\sqrt{2}} \overline{\ell}_L \ell_R + \frac{1}{\sqrt{2}} \overline{\ell}_L \ell_R (h+i c_\beta \sigma) + c^-_\beta \overline{\nu}_\ell \ell_R \sigma^+\right]\\
&+ \lambda^{e1}_H\left[ -\frac{v}{\sqrt{2}} \overline{\nu}_e N_1-\frac{1}{\sqrt{2}} \overline{\nu}_e N_1 (h-i c_\beta \sigma) + c^-_\beta \overline{e}_L N_1 \sigma^-\right] + (e 1\ra \mu 2,\tau 3)\\
&+ \lambda^{e\mu}_\phi \left[ \frac{v_{\phi}}{\sqrt{2}} \overline{e}_L \mu_R + \frac{1}{\sqrt{2}} \overline{e}_L \mu_R (\phi - i s_\beta c_\theta \sigma) - s^-_\beta \overline{\nu}_e \mu_R \sigma^+\right] + (e\mu \ra \tau e)\\
&+\lambda^{e 3}_\phi \left[ -\frac{v_{\phi}}{\sqrt{2}} \overline{\nu}_e N_3 - \frac{1}{\sqrt{2}}\overline{\nu}_e N_3 (\phi + i s_\beta c_\theta \sigma) - s^-_\beta \overline{e}_L N_3 \sigma^-\right]+ (e 3 \ra \mu 1)\\
&+ \lambda^{12}_S \left[ \frac{v_S}{\sqrt{2}} \overline{N}_1^c N_2 + \frac{1}{\sqrt{2}} \overline{N}_1^c N_2 (S+i s_\beta s_\theta\sigma)\right] \\
&+ \lambda^{13}_S \left[ \frac{v_S}{\sqrt{2}} \overline{N}_1^c N_3 + \frac{1}{\sqrt{2}} \overline{N}_1^c N_3 (S-i s_\beta s_\theta\sigma)\right] +\hc\,,
\end{split}
\label{eq:LFV_Yukawa}
\end{align}
which can be further simplified using $\bar\ell_L \ell_R (h+i c_\beta \sigma)+\hc = \bar\ell (h+i c_\beta \gamma_5 \sigma)\ell$, emphasizing the pseudoscalar nature of~$\sigma$.
The coupling to quarks is of the same form as in the Standard Model (since they are singlets under $U(1)'$):
\begin{align}
 -\L \ \supset \ \sum_{i,j=1,2,3} \lambda^{i j}_d \overline{Q}^i_L H d^j_R +\sum_{i,j=1,2,3} \lambda^{i j}_u \overline{Q}^i_L \tilde H u^j_R +\hc
\end{align}
Diagonalization of the mass matrices goes through as usual, via bi-unitary transformations; we end up with
\begin{align}
 -\L \ \supset\ \overline{\vec{d}}_L D_d \vec{d}_R H^0 + \overline{\vec{u}}_L D_u \vec{u}_R \overline{H}^0 + \overline{\vec{u}}_L V D_d \vec{d}_R H^+ - \overline{\vec{d}}_L V^\dagger D_u \vec{u}_R H^- +\hc\,,
\end{align}
with the matrices in generation space 
\begin{align}
D_d \equiv \sqrt{2}\, \mathrm{diag}(m_d,m_s,m_b)/v \,, && D_u \equiv \sqrt{2}\, \mathrm{diag}(m_u,m_c,m_t)/v \,,
\end{align}
and $V$ the usual unitary Cabibbo-Kobayashi-Maskawa matrix of the
Standard Model. In the SM, the terms with $H^\pm$ vanish in unitary
gauge, while in our case we have the Yukawa interactions:
\begin{align}
\begin{split}
 -\L \ \supset \ &\sum_i \frac{m^d_i}{v} \,\overline{d}^i_L d^i_R (h+i c_\beta \sigma) + \sum_i \frac{m^u_i}{v} \,\overline{u}^i_L u^i_R (h-i c_\beta \sigma)\\
&+c^-_\beta \frac{\sqrt{2}}{v} \sum_{i,j} m^d_j V_{ij} \,\overline{u}^i_L d^j_R \sigma^+
- c^-_\beta \frac{\sqrt{2}}{v} \sum_{i,j} m^u_j V^*_{j i} \,\overline{d}_L^i u^j_R \sigma^- +\hc
\end{split}
\end{align}
The flavor-changing interactions are suppressed by the Yukawa couplings $m_q/v$ and the angle $\cos \beta^- \simeq v_\phi /v$, compared to those induced by $W^\pm$. The interaction of the charged scalars with the quarks is very similar to the Two-Higgs-Doublet Model of Type I (2HDM-I)~\cite{2hdm1}, where the parameter $\cos\beta^-$ is denoted by $\tan\beta$. 
The corresponding bound $m_{\sigma^-} \gtrsim \unit[80]{GeV}$ on a charged scalar with decay channels $\sigma^- \ra \bar c s$, $\tau \bar\nu_\tau$, set by LEP~\cite{charged_higgs}, applies. Additional contributions from $\sigma^\pm$ to charged-current decays are already well suppressed; for example, the decay $\tau \ra \nu_\tau \sigma^- \ra \nu_\tau \overline{\nu}_\mu \mu$ has the width
\begin{align}
  \Gamma (\tau \ra \nu_\tau \overline{\nu}_\mu \mu) \equiv 	\Gamma_{\sigma^\pm} \simeq \frac{1}{192} \frac{1}{(2\pi)^3} \left(\frac{c^-_\beta}{m_{\sigma^-}}\right)^4  \left( \frac{m_\tau m_\mu}{v^2}\right)^2 m_\tau^5 \,,
\end{align}
resulting in an additional branching ratio of $\Gamma_{\sigma^\pm}/\Gamma_\mathrm{total} \simeq 10^{-11} (c^-_\beta)^4 (\unit[80]{GeV}/m_{\sigma^-})^4$, at least~$7$ orders of magnitude below the current sensitivity~\cite{PDG2010}. 

There are two different kinds of Lepton Family number Violation (LFV)
associated with this model, we will discuss them in the
following. Since we have chosen the charge $Y'=-1$ for our Higgs fields
$\phi$ and $S$, the $L_\mu-L_\tau$ number of a process will only
changed by one 
unit in the simplest Feynman diagrams, i.e.~we expect decays $\tau \ra e X$ and $\mu\ra e X$, but not $\tau \ra \mu X$.

\subsubsection{LFV Mediated by \texorpdfstring{$Z'$}{Z'}}
\label{sec:LFVbyZprime}
As can be seen immediately in Eq.~\eqref{eq:LFV_Yukawa}, the VEV
$v_\phi$ introduces nondiagonal elements in the mass matrix of the charged leptons:
\begin{align}
 	\mathcal{M}_\mathrm{leptons} =\frac{1}{\sqrt{2}}  \matrixx{ \lambda^{ee}_H v & \lambda^{e\mu}_\phi v_\phi & 0\\
0 & \lambda^{\mu\mu}_H v & 0\\
\lambda^{\tau e}_\phi v_\phi & 0 & \lambda^{\tau\tau}_H v} .
\label{eq:Mleptons}
\end{align}
The mass eigenbasis is obtained by means of a bi-unitary
diagonalization, i.e.~$\ell_L \ra U_L \ell_L$, $\ell_R \ra U_R
\ell_R$, with $U_L \neq U_R$. 
The relevant rotation matrices are 
\begin{align}
	U_L \simeq \matrixx{1 & \theta_{12}^L & 0 \\ -\theta_{12}^L & 1 & 0\\ 0 & 0 & 1} , &&
	U_R \simeq \matrixx{1 & 0 & \theta_{13}^R \\ 0 & 1 & 0 \\ -\theta_{13}^R & 0 & 1} ,
\end{align}
with $\theta_{12}^L \simeq \lambda^{e \mu}_\phi v_\phi / \sqrt{2}
m_\mu$ and $\theta_{13}^R \simeq \lambda^{\tau e}_\phi v_\phi / \sqrt{2} m_\tau$. 
Since these matrices operate in flavor
space, they do not change the normal $Z$-boson gauge interactions, but
the $Z'$ coupling:
\begin{align}
\begin{split}
 	j'_\mu Z'^\mu &= \sum_{i=L,R} \overline{\ell}_i \matrixx{0 & & \\ & 1 & \\ & & -1} \gamma_\mu \ell_i Z'^\mu\\
&\ra \sum_{i=L,R} \overline{\ell}_i U_i^\dagger \matrixx{0 & & \\ & 1 & \\ & & -1} U_i \gamma_\mu \ell_i Z'^\mu\,,
\end{split}
\end{align}
Since $U_i^\dagger \,\mathrm{diag}(0,1,-1) U_i$ is not diagonal, the
$Z'$ introduces interactions like $\tau \ra e Z'^* \ra e \mu\bar\mu$. 
This also generates a coupling of $Z'$ to electrons, suppressed by
$\theta_{L,R}^2$, 
and furthermore all the couplings become chiral, i.e.~the $Z'$ couples differently to left- and right-handed fermions.
The same reasoning applies to LFV mediated by neutral scalars, 
since they couple in a generation-dependent way as well. 
The $Z'$-mediated LFV decays are 
\begin{align}
\tau \ra e\, \overline \mu\, \mu\,, \qquad \tau \ra e\, \overline{\nu}_{\mu,\tau}\, \nu_{\mu,\tau}\,,\qquad \mathrm{ and }\qquad \mu \ra e \, \overline{\nu}_{\mu,\tau}\, \nu_{\mu,\tau}\,,
\label{eq:LFVprocesses}
\end{align}
the first of which can be probed in $B$-factories and leads to the constraint on $\theta_{13}^R$~\cite{belle}
\begin{align}
 \frac{\Gamma (\tau \ra e \mu\bar\mu)}{\Gamma (\tau \ra \mu \nu_\tau \bar\nu_\mu)} &\simeq (\theta_{13}^R)^2 \left( \frac{\unit[200]{GeV}}{M_{Z'}/g'}\right)^{4} \stackrel{!}{<} 1.6 \times 10^{-7} \,,
\end{align}
so for $v_\phi \sim m_\tau$ and $M_{Z'}/g' \sim \unit[200]{GeV}$ we find the bound $\lambda^{\tau e}_\phi < 10^{-3}$--$10^{-4}$ for the Yukawa coupling. The angle $\theta_{12}^L$ can not be probed in this way due to the challenging neutrinos in the final state. 
However, this angle contributes to the PMNS mixing matrix via the
charged current interactions, i.e.~$U_\mathrm{PMNS} = U_L^\dagger
U_\nu$, which most importantly adds to $s_{13}$ a term $\theta_{12}^L
s_{23}$, where $s_{23}$ denotes $\sin \theta_{23}$ of $U_\nu$. 
A relatively large $\theta_{13}$ can in consequence be generated without a strongly broken $\Uprime$, simply due to the interplay with the charged leptons (depending on the signs, a cancellation could occur as well). Since nothing in the motivation for our model depends on $\lambda^{e\mu}_\phi$ and $\lambda^{\tau e}_\phi$, we can make them arbitrarily small (and they can still be larger than the Yukawa coupling of the electron).
\subsubsection{LFV via Loops}
\begin{figure}[t]
\begin{center}
\includegraphics[scale=0.55]{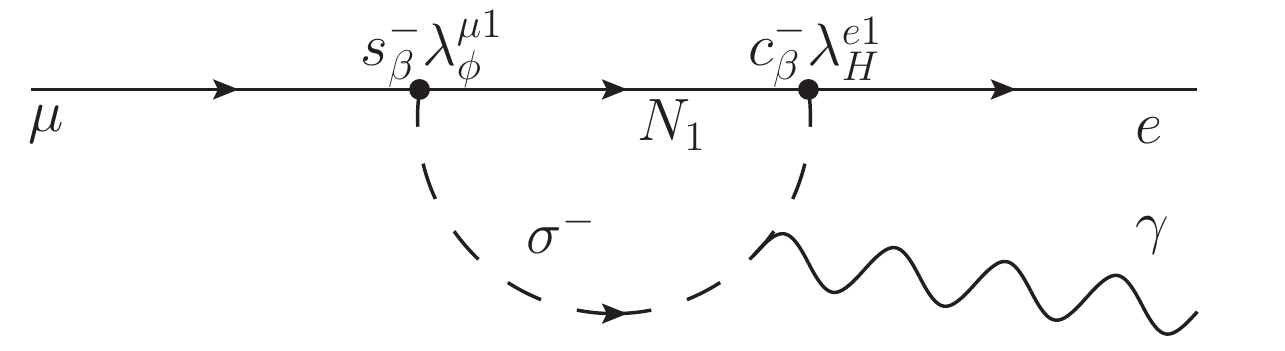}
\end{center}
\caption{\label{fig:LFV_loop}Charged-scalar mediated lepton-flavor-changing radiative decay.}
\end{figure}

The second source of LFV stems from the charged scalars, inducing the decays $\mu \ra e \gamma$ and $\tau \ra e \gamma$ via diagrams like Fig.~\ref{fig:LFV_loop}, with a heavy right-handed neutrino in the loop. Since these decays involve the same Yukawa coupling $\lambda^{\ell i}$ that generate the $U(1)'$-breaking elements in the neutrino mass matrix, they better not be too small in our model. Calculating the branching ratio of the decay $\mu\ra e\gamma$ in the approximation \mbox{$m_N \gg m_{\sigma^+},m_\mu,m_e$}, we find~\cite{Lavoura:2003xp}
\begin{align}
\begin{split}
 	\BR (\mu\ra e\gamma) &\simeq \frac{1}{192 \pi^2} \left( \frac{s_\beta^- \lambda_\phi^{\mu 1} c_\beta^- \lambda_H^{e 1} }{2 G_F m_N^2}\right)^2\simeq  \frac{\gamma^2}{96 \pi^2} \left( \frac{m_{\nu_e}^2/Y}{Y}\right)^2 \simeq 10^{-29}\, \gamma^2 \,,
\end{split}
\end{align}
which is highly suppressed by the heavy neutrino mass and poses no
bound on $\gamma = f/m_{\nu_e}$~(see Eq.~\eqref{eq:small_parameters}). We also see that the predicted LFV from the scalars is too low to be observed in any future experiment, as opposed to the $Z'$-mediated processes.

\subsubsection{Contribution to \texorpdfstring{$\Delta a_\mu$}{Delta amu}}
\begin{figure}[t]
\begin{center}
\includegraphics[height=0.30 \textwidth]{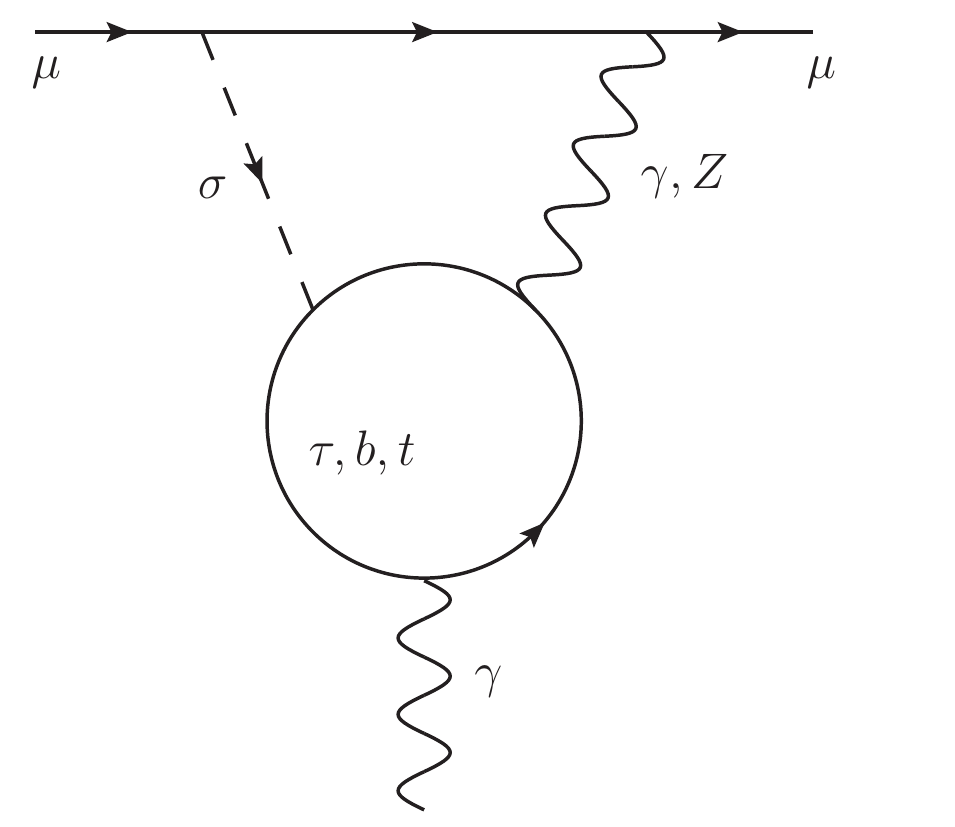}\hspace{4ex}
\includegraphics[height=0.30 \textwidth]{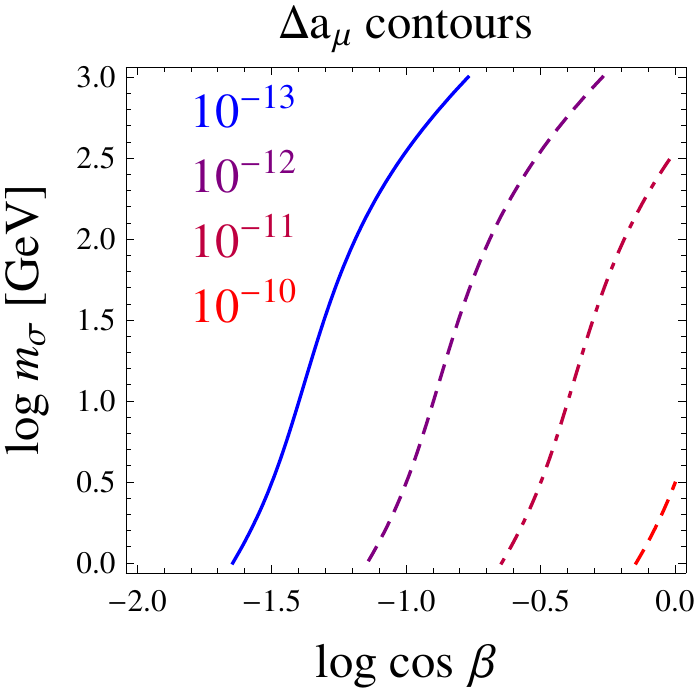}
\end{center}\vspace{-3ex}
\caption{Dominating two-loop Barr-Zee-type diagram contributing to $\Delta a_\mu$ (left), actual values for $\Delta a_\mu$ in the approximation $\cos\beta^- \simeq \cos\beta$, $m_\sigma \simeq m_{\sigma^-}$ (right).}
\label{fig:constraints_magnetic_moment_higgs}
\end{figure}

The physical scalars contribute to the anomalous magnetic moment of the muon via loop diagrams. Setting the LFV Yukawa couplings $\lambda^{e\mu}_\phi = \lambda^{\tau e}_\phi = 0$, only $h$, $\sigma$ and $\sigma^-$ couple directly to the muon. The one-loop contributions from the pseudoscalar $\sigma$ and the charged $\sigma^-$ are~\cite{zprimecontribution}
\begin{align}
 \Delta a_\mu^{1-\mathrm{loop}} = \frac{-m_\mu^4}{8 \pi^2 v^2} \int_0^1 \dd x \, \left[ 
\left( \frac{c_\beta^-}{m_{\sigma^-}} \right)^2 \frac{ x (1-x)}{1+ (x-1) m_\mu^2/m_{\sigma^-}^2} +
\left( \frac{c_\beta}{m_{\sigma}} \right)^2 \frac{ x^3}{1-x + x^2 m_\mu^2/m_{\sigma}^2}\right] ;
\end{align}
however, the two-loop contribution of $\sigma$ is also
important due to a larger coupling of $\sigma$ to
heavy fermions in the loop, which compensates the additional loop
suppression (see
Fig.~\ref{fig:constraints_magnetic_moment_higgs}). The dominant effect
gives \cite{Chang:2000ii}
\begin{align}
  \Delta a_\mu^{2-\mathrm{loop}} = \frac{\alpha}{8 \pi^3} \frac{m_\mu^2}{v^2} c_\beta^2 \sum_{f=t,b,\tau} N_\mathrm{color}^f Q_f^2 \frac{m_f^2}{m_\sigma^2} \int_0^1 \dd x \frac{\ln\left( \frac{m_f^2/m_\sigma^2}{x (1-x)}\right)}{m_f^2/m_\sigma^2 - x (1-x)}\,.
\end{align}
The combined one and two-loop contributions are shown in Fig.~\ref{fig:constraints_magnetic_moment_higgs}~(right) for the case $\cos\beta^-\simeq \cos\beta$, $m_\sigma\simeq m_{\sigma^-}$, corresponding to the $v_\phi\ll v$ limit we are interested in. As can be seen the effects are about~$2$ orders of magnitude too small to have any visible effect.

\subsection{Signatures at the LHC}
The effects of additional scalars in collider experiments, especially
concerning the disentanglement of different multi Higgs doublet
models, have been reviewed in Ref.~\cite{Barger:2009me}; since our
model is similar to the 2HDM-I in the decoupling limit, we expect
similar signatures. The best candidate for observation will be the
scalar $h$, with couplings reduced by mass mixing (which goes roughly
with $\mu/v$, of the same order as $\cos\beta$), which we did not
discuss before, and smaller branching ratios due to the additional
decay modes via the other scalars ($h\ra \sigma \sigma$, $ \sigma^+
\sigma^-$, $\phi\phi$, $\phi S$, $S S$) and in association with gauge
bosons ($h\ra Z W^-\sigma^+$, $A W^-\sigma^+$, $W^-\sigma^+$, $Z
\sigma$), most important for a heavy $h$. An analysis of the branching
ratios of $h$ will therefore not suffice to distinguish our model from
the 2HDM-I.

It is interesting to note that the $Z$--$Z'$ mixing angle goes roughly
quadratic in $v_\phi$ ($\xi \sim v_\phi^2/v_S^2$ from
Eq.~\eqref{eq:abc}), while the scalar mixing is linear ($\theta\sim
v_\phi/v_S$, $\cos \beta \sim \sin \theta$ from
Eq.~\eqref{eq:thetaandbeta}). This suggests better direct detection
prospects via Drell-Yan processes for the scalars than for $Z'$. 
Since the interactions of $\sigma,\sigma^\pm,\phi$ and $S$ with the
leptons are suppressed not only by $\cos\beta$, but also by their
small Yukawa couplings, whereas the gauge boson couplings scale with
$\cos\beta$, this sector will be the most interesting. For example the
decay channel $h\ra Z \sigma$ (discussed in
Ref.~\cite{deVisscher:2009zb}) scales with $\cos\beta$; the decay
$h\ra Z' \sigma$ is induced by mass mixing of the scalars and thus
goes roughly with $\mu/v$. This could lead to interesting signatures,
since the invariant mass of the subsequently created leptons gives
information about the virtual particles, their angular distribution
about the spin of the bosons and the rates of electrons, muons and
tauons and about the admixture of $Z'$ over $Z$. Such an analysis would however require a lot of luminosity.
In general, the most dominant effect of the scalars and $Z'$ will be
the difference in the $e,\mu,\tau$ rates due to $Z'$ decays.

We mention the obvious fact that a future muon collider would be the ideal experiment to test this model, basically in total analogy to $Z$ measurements at LEP. Since the $\Uprime$ symmetry in this model connects the heavy right-handed neutrino sector to the SM, this would also open up a way to probe for this special mechanism of neutrino mass generation. An analysis of these signatures lies outside the realm of this work, but has been performed for a similar model (based on gauged $B-L$ at the LHC) in Ref.~\cite{Basso:2008iv}.

\section{Extension to \texorpdfstring{$\boldsymbol{SU(2)_{L_\mu-L_\tau}}$}{SU(2)'}}
\label{sec:nonabelianextension}

Nonabelian family symmetries based on $SU(2)_H$ or $SU(3)_H$ (``horizontal symmetry'') have been discussed extensively in the literature~\cite{horizontal_symmetry}, although mainly with focus on the quark sector.
An extension from $\Uprime$ to a nonabelian group is natural since it
includes the electron into the symmetry. It also forces the kinetic
mixing angle $\chi$ to be zero at tree-level, because the field
strength tensor of the $SU(2)'$ gauge bosons is not a gauge-invariant
object. 
Reference~\cite{joshi} also contains discussions of an $SU(2)$
extension of $\Uprime$, but with no emphasis on the
neutrino structure. 
Constructing a three-dimensional representation with a diagonal generator $T_\mathrm{diag}^{SU(N)} (\mu) = - T_\mathrm{diag}^{SU(N)} (\tau)$ and $T_\mathrm{diag}^{SU(N)} (e)=0$ is possible for $N=2$ and $N=3$. Since $SU(3)$ can be seen as an extension of $SU(2)$ we will not consider it in the following. The extension from $\Uprime$ to $SU(2)'$ remains anomaly-free even without right-handed neutrinos, partly because the $SU(2)$ only has real and pseudoreal representations.\footnote{The only possible anomaly is $SU(2)'$-$SU(2)'$-$ U(1)_Y$, which vanishes as long as the charged leptons are in the same $SU(2)'$ representation.} For the $SU(2)'$, we have two possibilities concerning the representation of electron, muon and tauon:
\begin{enumerate}[(i)]
	\item irreducible: $e$, $\mu$ and $\tau$ form an $SU(2)'$ triplet,
	\item reducible: $e$ transforms as a singlet and $(\mu,\tau)$ form a doublet.
\end{enumerate}
The latter case once again treats the electron differently than muon
and tauon, furthermore it is not possible to implement a seesaw-I
mechanism, so we discuss it only briefly in Appendix
\ref{app:reduciblerep}. 
In the following we will therefore use the $G_\sm \times SU(2)'$ representations 
\begin{align}
	\vec{L}\equiv (L_\mu,\, L_e,\, L_\tau) \sim (\vec{1},\vec{2},-1)(\vec{3})\,, &&
	\vec{\ell}_R \equiv (\mu_R,\,e_R,\,\tau_R) \sim (\vec{1},\vec{1},-2)(\vec{3})\,.
	\label{eq:irreduciblerep}
\end{align}
We will also refer to the $SU(2)'$ as ``leptospin" for convenience later on. Since $\Uprime$ is the $SU(2)'$ subgroup generated by
\begin{align}
	T_z^{(3)} = \matrixx{1 & 0 & 0\\ 0 & 0 & 0\\ 0 & 0 & -1} ,
	\label{eq:diagonalgenerator}
\end{align}
we expect a possible breaking pattern $SU(2)' \ra \Uprime \ra$~nothing, which might still resolve $\Delta a_\mu$ and explain the neutrino mixing angles. In the next sections we will comment on the difficulties concerning this task.

It proves convenient for the most part to work in the flavor basis $(\mu,e,\tau)$, as already used in Eq.~\eqref{eq:irreduciblerep} and~\eqref{eq:diagonalgenerator}; however, to make the neutrino mass matrices look more familiar, the transformation back to the usual $(e,\mu,\tau)$ basis can be performed via 
\begin{align}
	\vec{L} \ra U \vec{L}\,, &&
	\mathcal{M} \ra U \mathcal{M} U\,, &&
		U = \matrixx{0 & 1 & 0\\ 1 & 0 & 0\\ 0 & 0 & 1} ,
\end{align}
where the matrix $U$ satisfies $U= U^{-1} = U^T$.

\subsection{Lepton Masses}
The allowed mass terms for the charged leptons are generated by
\begin{align}
	\L\ \supset\  Y_H  \vec{\overline{L}} H \vec{\ell}_R\,,
\end{align}
which gives $m_e = m_\mu = m_\tau$. To break this symmetry we introduce an $SU(2)'$ triplet $\Delta$ and a pentet (leptospin-2) $\Sigma$, with the same $G_\sm = SU(3)_C \times SU(2)_L \times U(1)_Y$ quantum numbers as the standard Higgs $H$, i.e.:
\begin{align}
	H \sim (\vec{1},\vec{2},+1)(\vec{1})\,, && \Delta \sim (\vec{1},\vec{2},+1)(\vec{3})\,, && \Sigma \sim (\vec{1},\vec{2},+1)(\vec{5})\,.
\end{align}
In matrix notation, we have (see App.~\ref{app:fieldtrafos} for a short collection of used representations)
\begin{align}
	\Delta = \frac{1}{\sqrt{2}} \matrixx{\Delta^0 & \Delta^+ & 0 \\ \Delta^- & 0 & \Delta^+ \\ 0 & \Delta^- & -\Delta^0} , &&
	\Sigma = \frac{1}{\sqrt{6}} \matrixx{\Sigma^0 & \sqrt{3}\, \Sigma^+ & \sqrt{6}\, \Sigma^{++} \\ \sqrt{3}\, \Sigma^- & -2\, \Sigma^0 & -\sqrt{3}\, \Sigma^+ \\ \sqrt{6} \, \Sigma^{--} & -\sqrt{3}\, \Sigma^- & \Sigma^0} ,
\end{align}
where the superscript denotes the $L_\mu - L_\tau$ charge of the $SU(2)_L$ doublet, not the electric charge. In fact, all of the following discussion is focused on flavor space, the $SU(2)_L$ contractions will not be used. 
Since the two leptospin-1 fields $\vec{L}$ and $\vec{\ell}_R$ can couple to leptospin-0, 1 and 2, the most general allowed Yukawa couplings are given by
\begin{align}
	\L \ \supset\   \vec{\overline{L}} \left(Y_H \,H +Y_\Delta\, \Delta + Y_\Sigma\, \Sigma\right) \vec{\ell}_R\,,
\end{align}
so if the fields acquire VEVs that leave $\Uprime$ intact (i.e.~only $\Delta^0$ and $\Sigma^0$), we get the masses
\begin{align}
\begin{split}
	m_\mu &= Y_H \, \langle H\rangle +Y_\Delta \langle \Delta^0 \rangle / \sqrt{2}+ Y_\Sigma \langle \Sigma^0 \rangle / \sqrt{6} \,, \\
	m_e &= Y_H \, \langle H\rangle -2 Y_\Sigma \langle \Sigma^0 \rangle / \sqrt{6} \,, \\
	m_\tau &= Y_H \, \langle H\rangle - Y_\Delta \langle \Delta^0 \rangle / \sqrt{2}+ Y_\Sigma \langle \Sigma^0 \rangle / \sqrt{6} \,.
\end{split}
\end{align}
To get the charged-lepton masses right we need all three VEVs, the small electron mass is the result of a fine-tuned cancellation. Specifically, we have
\begin{align}
\begin{split}
	Y_H \, \langle H\rangle &=(m_\mu+m_e + m_\tau)/3\, \simeq\, \unit[0.6]{GeV}\,,\\
	Y_\Delta \langle \Delta^0 \rangle &=  (m_\mu - m_\tau)/ \sqrt{2}\, \simeq \, -\unit[1.2]{GeV}\,,\\
	Y_\Sigma \langle \Sigma^0 \rangle &= (m_\mu - 2\, m_e + m_\tau)/\sqrt{6}\, \simeq \, \unit[0.8]{GeV}\,.
\end{split}
\label{eq:yukawaschargedleptons}
\end{align}
Since all these $SU(2)_L$ doublets contribute to $M_W$ and $M_Z$, we have the boundary condition $\langle H\rangle^2  +\langle \Delta^0 \rangle^2 + \langle \Sigma^0 \rangle^2 \simeq (\unit[174]{GeV})^2$, and because $\langle H\rangle$ gives the mass to the top-quark, it will be the largest of these three VEVs; for approximations, we will use $\langle \Delta^0 \rangle,\langle \Sigma^0 \rangle \sim \mathcal{O}(10)\unit{GeV}$. 
Seeing that this breaking scheme leaves $\Uprime$ as an exact symmetry, there will not be any mixing of $Z$ with the $SU(2)'$ gauge bosons $X_i$ at tree-level.
The kinetic terms for the charged leptons obviously lead to LFV:
\begin{align}
	\L \ \supset \ i  \vec{\overline{L}} \slashed{D} \vec{L} + i  \vec{\overline{\ell}}_R \slashed{D} \vec{\ell}_R\,,
\end{align}
with covariant derivative $D_\mu = \del_\mu -i g' X_j T^{(3)}_j$. It proves convenient to define the two gauge fields $X^\pm \equiv (X_1 \mp i X_2)/\sqrt{2}$, the $SU(2)'$ gauge interactions then take the form
\begin{align}
	\L/g'\ \supset \ \overline{\mu} \slashed{X}_3 \mu - \overline{\tau} \slashed{X}_3 \tau
	+\overline{e} \slashed{X}^+ \tau +\overline{e} \slashed{X}^- \mu +\overline{\mu} \slashed{X}^+ e +\overline{\tau} \slashed{X}^- e\,,
\end{align}
which generate the process $\tau \ra e \, \overline{\mu} \, e$ at tree-level (plus other, less constrained decays involving neutrinos). The branching ratio for this process is less than $1.5 \times 10^{-8}$~\cite{PDG2010}, leading to a constraint
\begin{align}
M_{X^\pm}/g' \gtrsim \unit[\mathcal{O}(10)]{TeV}\,.
\label{eq:lfvconstraint}
\end{align}
Since such a high breaking scale can not be realized with $SU(2)_L$
doublets, it is necessary to introduce more scalar fields that break
$SU(2)'$ but do not contribute to $M_Z$ and $M_{W^\pm}$. Fortunately,
this fits into the neutrino mass generation via seesaw. Once again we
will use the breaking scale to find the $N_R$-scale, in analogy to
Section \ref{sec:neutrino_sector}.

\subsubsection{Majorana Masses}
Introducing right-handed neutrinos $\vec{N}\equiv (N_\mu, N_e,
N_\tau)\sim (\vec{1},\vec{1},0)(\vec{3})$, conveniently written as 
\begin{align}
	N = \matrixx{N_e/\sqrt{2} & N_\mu \\ N_\tau & -N_e/\sqrt{2}} ,
\end{align}
allows for the $SU(2)'$-invariant mass term $\tr (\overline{N}^c N) = \overline N_e^c N_e + \overline N_\mu^c N_\tau + \overline N_\tau^c N_\mu$, leading to a Majorana mass matrix
\begin{align}
	\mathcal{M}_R = m_R \matrixx{0 & 0 & 1 \\ 0 & 1 & 0 \\ 1 & 0 & 0} .
	\label{eq:majoranamass}
\end{align}
Note that the eigenvalues of $\mathcal{M}_R$ are degenerate. 
As far as the allowed Yukawa couplings go, the coupling of the symmetric bilinear $\overline{N}^c_i N_j$ to a leptospin-1 field vanishes,\footnote{The coupling of three leptospin-1 fields uses the $SU(2)'$ invariant antisymmetric symbol $\epsilon_{ijk}$.} so we introduce another leptospin-2 field $\Omega \sim (\vec{1},\vec{1},0)(\vec{5})$, transforming as a singlet under $G_\sm$. Since it carries no other quantum numbers, we can choose the fields real, i.e.~$\Omega$ is an hermitian matrix:
\begin{align}
	 \Omega = \frac{1}{\sqrt{6}} \matrixx{\Omega^0 & \sqrt{3}\, \Omega^+ & \sqrt{6}\, \Omega^{++} \\ \sqrt{3}\, \Omega^- & -2\, \Omega^0 & -\sqrt{3}\, \Omega^+ \\ \sqrt{6} \, \Omega^{--} & -\sqrt{3}\, \Omega^- & \Omega^0} && \mathrm{ with } && (\Omega^-)^\dagger = \Omega^+, \ (\Omega^{--})^\dagger = \Omega^{++}\,.
\end{align}
This allows for the Yukawa terms
\begin{align}
\begin{split}
	\L \ &\supset \  \frac{Y_\Omega}{2 \sqrt{6}} \  \tr \left[ \matrixx{\overline{N}^c_e & \overline{N}^c_\mu & 0 \\  \overline{N}^c_\tau & 0 & \overline{N}^c_\mu \\ 0 &  \overline{N}^c_\tau & -\overline{N}^c_e} 
	  \matrixx{\Omega^0 & \sqrt{3}\, \Omega^+ & \sqrt{6}\, \Omega^{++} \\ \sqrt{3}\, \Omega^- & -2\, \Omega^0 & -\sqrt{3}\, \Omega^+ \\ \sqrt{6} \, \Omega^{--} & -\sqrt{3}\, \Omega^- & \Omega^0}
	  \matrixx{{N}_e & {N}_\mu & 0 \\  {N}_\tau & 0 & {N}_\mu \\ 0 &  {N}_\tau & -{N}_e} \right]\\
	  &= \frac{Y_\Omega}{2 \sqrt{6}} \   \matrixx{-\overline{N}^c_\mu & \overline{N}^c_e & \overline{N}^c_\tau}
	  \matrixx{\sqrt{6}\,\Omega^{--} & -\sqrt{3}\, \Omega^- & \Omega^0\\ -\sqrt{3}\, \Omega^- & 2\, \Omega^0 & \sqrt{3}\, \Omega^+\\ \Omega^0 & \sqrt{3}\, \Omega^+ & \sqrt{6}\,\Omega^{++}}
	  \matrixx{-N_\mu \\ N_e \\ N_\tau} ,
\end{split}
\end{align} 
where the first line shows explicitly the gauge invariance and the second line the symmetric nature of the coupling. A nonzero VEV $\langle \Omega^0 \rangle$ can be used to break the degeneracy of the $N_e$ and $N_{\mu,\tau}$ masses and lead to the general $\Uprime$ invariant Majorana mass matrix~\eqref{eq:mutaumajoranamatrix}.

\subsubsection{Dirac Neutrino Masses}
In direct analogy to the charged-lepton masses we have
\begin{align}
		\L \supset  \vec{\overline{L}} \left(\tilde Y_H \, \tilde H +\tilde Y_\Delta\, \tilde \Delta + \tilde Y_\Sigma\,\tilde  \Sigma\right) \vec{N}_R\,,
\end{align}
which leads to a diagonal Dirac-matrix $m_D$ with nondegenerate eigenvalues after $ \Delta^0 \ra \langle \Delta^0 \rangle$, $ \Sigma^0 \ra \langle \Sigma^0 \rangle$. For the definition of the tilde-fields see App.~\ref{app:fieldtrafos}.

\subsection{Masses for the Gauge Bosons}
The Lagrangian for the $SU(2)'$-charged scalars:
\begin{align}
	\L \ \supset \  \tr \left( (D_\mu \Delta )^\dagger D^\mu \Delta\right) +\tr \left( (D_\mu \Sigma )^\dagger D^\mu \Sigma\right)+ \frac{1}{2}\, \tr \left( (D_\mu \Omega )^\dagger D^\mu \Omega \right)\,,
\end{align}
results in mass-terms for $X_i$ after $SU(2)'$-breaking via $\Delta^0$, $\Sigma^0$ and $\Omega^0$:
\begin{align}
	M^2_{X^\pm} =2\, g'^2 \langle \Delta^0\rangle^2 + 6 \,g'^2 \langle \Sigma^0 \rangle^2 + 3 \,g'^2 \langle \Omega^0 \rangle^2\,, && M^2_{X_3} = 0\,.
\end{align}
Because of the constraint~\eqref{eq:lfvconstraint}, the VEV $ \langle \Omega^0 \rangle$ should be around $\unit[10]{TeV}$, which is fine for an $N_R$-scale around $\unit[100]{TeV}$. We mention that $M_{X_3}$ can be pushed arbitrarily high via the VEV of an $SU(2)'$ doublet, without affecting any of the discussed lepton phenomenology.

\subsection{Scalar Potential}
\label{sec:scalar_potential}
One $SU(2)'$ singlet $(H)$, one triplet $(\Delta)$ and two leptospin-2 fields ($\Sigma$ and $\Omega$) result in a gauge-invariant potential $V (H,\Delta,\Sigma,\Omega) = V_2 + V_3 + V_4$ with
\begin{align}
\begin{split}
	V_2 &= 
\mu_H^2 \, H^\dagger H+ \mu_\Delta^2 \tr ( \Delta^\dagger \Delta) + \mu_\Sigma^2 \tr ( \Sigma^\dagger \Sigma)+ \mu_\Omega^2 \tr ( \Omega \Omega)\,,\\
		V_3 &=  
+ \mu_1 H \tr (\Sigma^\dagger \Omega) 	+\mu_2  \tr (\Delta^\dagger \Omega \Delta)  +\mu_3  \tr (\Sigma^\dagger \Omega \Delta)\\
			&\quad +\mu_4  \tr (\Sigma^\dagger \Sigma \Omega)	+\mu_5 \tr (\Omega \Omega \Omega)  +\hc\,,
\end{split}
\label{eq:V2andV3}
\end{align}
and finally some of the quartic interactions:
\begin{align}
\begin{split}
	V_4 &= \lambda_H (H^\dagger H)^2 
	+ \lambda_\Delta\, (\tr( \Delta^\dagger \Delta))^2  
	+ \lambda_\Sigma\, (\tr (\Sigma^\dagger \Sigma))^2 
	+ \lambda_\Omega\, (\tr (\Omega \Omega))^2 
	+\lambda_1 \det (\Delta^\dagger \Delta ) \\
	&\quad +\lambda_2 \tr (\Sigma^\dagger \Sigma \Sigma^\dagger \Sigma) 
	+ \lambda_3 \tr (\Sigma^\dagger \Sigma^\dagger \Sigma\Sigma)
	+\lambda_{4} H \tr (\Sigma^\dagger\Omega \Omega) 
	+\lambda_{5} H \tr (\Delta^\dagger\Omega \Omega) \\
	&\quad + \lambda_6 H \tr (\Delta^\dagger \Sigma^\dagger \Sigma) 
	+ \lambda_7 H \tr (\Delta^\dagger \Delta^\dagger \Delta) 
	+ \lambda_8 H \tr ( \Sigma^\dagger \Sigma^\dagger \Sigma) 
	+ \lambda_9 H \tr (\Delta^\dagger \Delta^\dagger \Sigma)\\
	&\quad + \lambda_{10} |H|^2 \tr ( \Delta^\dagger \Delta)  
	+ \lambda_{11} |H|^2 \tr ( \Sigma^\dagger \Sigma) 
	+ \lambda_{12} |H|^2 \tr ( \Omega \Omega)  
	+ \lambda_{13}  \tr ( \Delta^\dagger \Delta) \tr ( \Sigma^\dagger \Sigma) \\
	&\quad + \lambda_{14}  \tr ( \Delta^\dagger \Delta) \tr ( \Omega \Omega)
	+  \lambda_{15} \tr ( \Sigma^\dagger \Sigma) \tr ( \Omega \Omega) 
	+\lambda_{16}  \tr (\Sigma^\dagger \Sigma \Omega \Omega) 
	+\lambda_{17}  \tr (\Delta^\dagger \Delta \Omega \Omega) \\
	&\quad +\lambda_{18} \tr (\Delta^\dagger \Delta \Sigma^\dagger \Sigma)
	+\lambda_{19} \tr (\Delta^\dagger \Sigma \Omega \Omega)
	+\lambda_{20} \tr (\Delta^\dagger \Sigma \Delta^\dagger \Delta)
	+\lambda_{21} \tr (\Delta^\dagger \Sigma \Sigma^\dagger \Sigma) \\
	& \quad + \lambda_{22} \tr (\Sigma \Sigma) \,\tr (\Delta^\dagger \Delta^\dagger)
	+\lambda_{23} \tr (\Sigma \Sigma \Delta^\dagger \Delta^\dagger)
	+\ldots +\hc
\end{split}
\label{eq:V4}
\end{align}
The potential is obviously very complicated to analyze, so we will only discuss the potential for $\Omega$:
\begin{align}
	V (\Omega) =  \mu_\Omega^2 \tr ( \Omega \Omega) +\mu_5 \tr (\Omega \Omega \Omega) 
+ \lambda_\Omega\, (\tr (\Omega \Omega))^2\,,
\end{align}
It can be shown that one can eliminate either $\Omega^{++}$ or $\Omega^+$ via $SU(2)'$ gauge transformations (for hermitian $\Omega$), while making the other fields real (App.~\ref{app:killtrafo}). One can therefore study $V (\Omega)$ as a function of the two real parameters $\Omega^0$ and $\re \Omega^+$. For $\mu_\Omega^2 < 0$, the potential has a minimum at
\begin{align}
	\langle \Omega^0 \rangle = 
\sqrt{\frac{-\mu_\Omega^2}{2\lambda_\Omega}} + \frac{3 }{8\sqrt{6}}\,\frac{\mu_5}{\lambda_\Omega} + \mathcal{O}(\mu_5^2/\mu_\Omega) \,, && \langle \Omega^+ \rangle = 0\,,
\end{align}
where we assumed $0< \mu_5 \ll \mu_\Omega$ and a positive VEV. In
unitary gauge ($\Omega^\pm$ is eaten by $X^\pm$), $\Omega^0$ receives
a mass $M_{\Omega^0}^2 \sim \lambda\, \langle \Omega^0 \rangle^2$ while $\Omega^{++}$ is
comparatively light, $M_{\Omega^{++}}^2 \sim \mu_5\,  \langle \Omega^0 \rangle$.

The above discussion was meant to show the possibility of the
aforementioned breakdown $SU(2)'\ra \Uprime\ra$~nothing, which can
still accommodate the nice features of the pure $\Uprime$ model,
namely a motivation for the maximal atmospheric mixing
angle~$\theta_{23}$ and the resolution of the magnetic moment of the
muon. To complete the model, i.e.~break $\Uprime$, one would need to
examine the full scalar potential~(\ref{eq:V2andV3}, \ref{eq:V4}), a
task that goes beyond the scope of this paper. We merely point out
that the required VEVs of the $\Uprime$ charged scalars need to be
such that the off-diagonal parts in the neutrino mass matrix $m_D$ are
large enough to generate a viable $U_\mathrm{PMNS}$ mixing matrix,
while the off-diagonal charged-lepton entries need to be small enough
to allow a $Z'$ with $M_{Z'}/g'\sim \unit[200]{GeV}$ without large LFV.

\section{Conclusion}
\label{sec:conclusion}
We constructed a viable extension of the Standard Model based on an
additional gauge group $\Uprime$. We discussed the most general
low-energy Lagrangian for a broken $\Uprime$, including mixing effects
with the $Z$-boson, and identified the parameter space allowed by
electroweak precision measurements. The goodness-of-fit can be
improved significantly with a $Z'$ at the electroweak scale, mostly
due to the resolved anomaly of the muons magnetic moment. As a side
effect of the nonuniversal gauge coupling, nonstandard neutrino
interactions are induced, potentially testable by future neutrino
oscillation experiments.  To complete the model we introduced an
economic scalar field sector that breaks the additional gauge symmetry spontaneously,
generating a viable neutrino mass matrix at tree-level, which features
nearly maximal mixing in the atmospheric sector and nonzero~$\theta_{13}$. Neutrino masses are expected to be quasi- or partially
degenerate and lead to testable
neutrinoless double $\beta$-decay, whereas the heavy right-handed
neutrinos are light enough to be produced at a future muon collider
via $Z'$ gauge interactions. 
The scalar sector of the theory is similar to other two-Higgs-doublet
models, introducing a small mixing between the physical
scalars. $Z'$-mediated lepton family number violation can be tested in
upcoming experiments and distinguishes this model from others via its
selected allowed modes.

The nonabelian extension of $\Uprime$ to $SU(2)'$ naturally includes
the electron into the symmetry and allows for a breakdown that leaves
$\Uprime$ exact at the electroweak scale, maintaining the nice
features of the pure $\Uprime$ model.

\begin{acknowledgments}
We thank Jens Erler for providing us with a recent version of GAPP. 
This work was supported by the ERC under the Starting Grant 
MANITOP and by the DFG in the Transregio 27. JH acknowledges support by the IMPRS-PTFS.
\end{acknowledgments} 

\appendix

\section{Field Transformations and Representations}
\label{app:fieldtrafos}
A field $\varphi$ in a particular representation of the gauge group $G_\sm \times \Uprime$ or $G_\sm \times SU(2)'$ is specified by the numbers 
\begin{align}
\begin{split}
	&\left(\vec{\dim}\, R_{SU(3)}(\varphi),\, \vec{\dim}\, R_{SU(2)} (\varphi),\, Y (\varphi) \right) \left( L_\mu (\varphi)-L_\tau (\varphi) \right) \quad \mathrm{ or }\\
	&\left(\vec{\dim}\, R_{SU(3)}(\varphi),\, \vec{\dim}\, R_{SU(2)} (\varphi),\, Y (\varphi) \right) \left( \vec{\dim}\, R_{SU(2)'} (\varphi) \right)\,,
\end{split}
\end{align}
respectively, where $Y (\varphi) = 2\, Q (\varphi) - 2\, T_z^{SU(2)} (\varphi)$ denotes the hypercharge. To distinguish more easily between charges and dimensions of representations, the dimensions are set in boldface.

\subsection{\texorpdfstring{$\boldsymbol{SU(2)'}$}{SU(2)'} Representations}
The $SU(2)'$-triplet $\Delta \sim (\vec{1},\vec{2},+1)(\vec{3})$ and pentet $\Sigma \sim (\vec{1},\vec{2},+1)(\vec{5})$ can be written as vectors like
\begin{align}
	\Delta = \matrixx{-\Delta^+, & \Delta^0,  & \Delta^-}^T , &&
	\Sigma = \matrixx{\Sigma^{++}, & -\Sigma^+, & \Sigma^0, & \Sigma^-, & \Sigma^{--}}^T ,
	\label{eq:vectorrep}
\end{align}
which transform like $\Delta \ra U \Delta$, $\Sigma \ra U\Sigma$ with $U = \exp ( -i \theta_j T_j^{(s)} )$ and the $SU(2)'$-generators for the $s$-dimensional representation $\vec{T}^{(s)}$, explicitly:
\begin{align}
	T^{(3)}_x &= \frac{1}{\sqrt{2}} \matrixx{0 & 1 & 0 \\ 1 & 0 & 1 \\ 0 & 1 & 0} , & T^{(5)}_x &= \matrixx{0 & 1 & 0 & 0 & 0 \\ 1 & 0 & \sqrt{3/2} & 0 & 0\\ 0 & \sqrt{3/2} & 0 & \sqrt{3/2} & 0 \\ 0 & 0 & \sqrt{3/2} & 0 & 1 \\ 0 & 0 & 0 & 1 & 0} ,\\
	T^{(3)}_y &= \frac{1}{\sqrt{2}} \matrixx{0 & -i & 0 \\ i & 0 & -i \\ 0 & i & 0} , & T^{(5)}_y &= \matrixx{0 & -i & 0 & 0 & 0 \\ i & 0 & -i\sqrt{3/2} & 0 & 0\\ 0 & i\sqrt{3/2} & 0 & -i\sqrt{3/2} & 0 \\ 0 & 0 &i \sqrt{3/2} & 0 & -i \\ 0 & 0 & 0 & i & 0} , \\
	T^{(3)}_z &= \matrixx{1 & 0 & 0 \\ 0 & 0 & 0 \\ 0 & 0 & -1} ,&
	T^{(5)}_z &= \matrixx{2 & 0 & 0 & 0 & 0 \\ 0 & 1 & 0 & 0 & 0\\ 0 & 0 & 0 &0 & 0 \\ 0 & 0 & 0 & -1 & 0 \\ 0 & 0 & 0 & 0 & -2} .
\end{align}
A more convenient representation is given by $3\times 3$-matrices transforming like $M \ra U M U^\dagger$, which can be obtained with the help of Clebsch-Gordan-coefficients:
\begin{align}
	\Delta = \frac{1}{\sqrt{2}} \matrixx{\Delta^0 & \Delta^+ & 0 \\ \Delta^- & 0 & \Delta^+ \\ 0 & \Delta^- & -\Delta^0} , &&
	\Sigma = \frac{1}{\sqrt{6}} \matrixx{\Sigma^0 & \sqrt{3}\, \Sigma^+ & \sqrt{6}\, \Sigma^{++} \\ \sqrt{3}\, \Sigma^- & -2\, \Sigma^0 & -\sqrt{3}\, \Sigma^+ \\ \sqrt{6} \, \Sigma^{--} & -\sqrt{3}\, \Sigma^- & \Sigma^0} ,
\end{align}
where, as before, the superscript denotes the $\Uprime$ charge of the field. The leptospin-1 field also has a representation as a $2\times 2$-matrix:
\begin{align}
	\Delta = \matrixx{\Delta^0/\sqrt{2} & \Delta^+\\ \Delta^- & -\Delta^0/\sqrt{2}} .
\end{align}
The weird sign in Eq.~\eqref{eq:vectorrep} was chosen to make the matrix representations of $\Delta$ and $\Sigma$ more symmetric, we could of course redefine $\Delta^+ \ra - \Delta^+$ to shift the sign to the matrices. The correct mapping between representations is only important when using both in the same Lagrangian to build invariants, e.g.~to show the equality
\begin{align}
	\tr \left( \Delta_{3\times 3}^\dagger\, \Sigma\, \Delta_{3\times 3} \right) = -\frac{1}{2} \matrixx{-\Delta^+, & \Delta^0, & \Delta^-}^* \, \Sigma\,  \matrixx{-\Delta^+\\ \Delta^0 \\ \Delta^-}  .
\end{align}
Other useful identities to build the scalar potential:
\begin{align}
	\left( \tr\, \Delta^\dagger_{3\times 3}\, \Delta_{3\times 3}\right)^2 &= 2  \, \tr\left( \Delta^\dagger_{3\times 3}\,  \Delta_{3\times 3}\, \Delta^\dagger_{3\times 3}\,  \Delta_{3\times 3}\right)\,,\\
	&= \tr\left(  \Delta^\dagger_{2\times 2} \,  \Delta_{2\times 2} \, \Delta^\dagger_{2\times 2} \,  \Delta_{2\times 2}\right) + 2 \, \det \left( \Delta^\dagger_{2\times 2} \,  \Delta_{2\times 2}\right)\,.
\end{align}
One can also show that the invariants $\tr \Delta^\dagger \, \Delta^\dagger \Delta \Delta$ and $\tr \Delta^\dagger \Delta^\dagger \tr \Delta \Delta$ can be expressed via $\tr\left(  \Delta^\dagger_{2\times 2} \,  \Delta_{2\times 2} \, \Delta^\dagger_{2\times 2} \,  \Delta_{2\times 2}\right)$ and $\det \left( \Delta^\dagger_{2\times 2} \,  \Delta_{2\times 2}\right)$.

To form Yukawa couplings with the right-handed neutrinos, it is convenient to define an $SU(2)_L$ doublet with opposite hypercharge, e.g.~via $\tilde{H} = -i \sigma_2 H^*$. For the nontrivial $SU(2)'$ fields $\Delta$ and $\Sigma$, the corresponding definition is
\begin{align}
	\tilde{\Delta} \equiv -i \sigma_2 \, \epsilon' \Delta^* \epsilon' \sim (\vec{1},\vec{2},-1)(\vec{3})\,,
\end{align}
where $\epsilon'$ acts on the $SU(2)'$ indices and takes the form
\begin{align}
	\epsilon' \propto \matrixx{0 & 0 & 1\\ 0 & -1 & 0\\ 1 & 0 & 0} ,
\end{align}
when $\Delta$ (or $\Sigma$) is written as a $3\times 3$-matrix.

\subsection{Elimination of \texorpdfstring{$\boldsymbol{\Omega^{++}}$}{Omegapp} from \texorpdfstring{$\boldsymbol{\langle \Omega \rangle}$}{<Omega>}}
\label{app:killtrafo}
In the discussion of the vacuum structure of the potential $V(\Omega)$ in Sec.~\ref{sec:scalar_potential} we made use of the fact that a VEV of $\Omega^{++}$ can be rotated away via $SU(2)'$ transformations. We will now briefly proof this claim. We decompose the complex fields $\Omega^{++}$ and $\Omega^+$ into real and imaginary parts. It is clear that a $z$-transformation can be used to make $\Omega^{++}$ real, so a general $SU(2)'$ transformation takes the form
\begin{align}
	\exp\left(-i \tilde z\, 	T^{(5)}_z\right) \,\exp\left(-i y\, 	T^{(5)}_y\right) \, \exp\left( -i  z \,	T^{(5)}_z\right) \, \matrixx{a, & - (b + i c), & d, & b-i c, &  a}^T .
\end{align}
Ignoring the $\tilde z$ transformation for now, the demand for a vanishing first component of the above vector takes the form (split into real and imaginary part):
\begin{align}
	\cos y \, ( -2 a \, \cos z \,  \sin z) + \sin y \,  (c \cos z - b \sin z) &= 0\,,\\
	a (3 + \cos 2 y)\,  \cos 2 z + \sqrt{6} \,  d \, \sin^2 y + 2 \sin 2 y \, (b \cos z + c \sin z) &=0\,.
\end{align}
The first equation can be readily solved for given $z$, so we plug the solution into the second equation to obtain
\begin{align}
	- 4 b^2 + 4 c^2 + \sqrt{6} \,  a d + \left( a^2 + 4 ( b^2 + c^2)\right) \cos 2 z - a \left( \sqrt{6} \,  d \cos 4 z + a \cos 6 z\right) = 0\,,
\end{align}
which can be shown to have real solutions by expressing $\cos n z$ through $\tan z \equiv t$:
\begin{align}
	f (t) \equiv c^2 +  \left( 2 a^2 - b^2 + 2 c^2 + \sqrt{6} \,  a d\right) t^2 + \left( c^2 - 2 a^2 - 2 b^2 + \sqrt{6} \,  a d\right) t^4 - b^2 t^6 \stackrel{!}{=} 0\,.  
\end{align}
$f(t)$ has real zeros because $f(0)>0$ and $f (t\ra \infty) < 0$. Hence we always find $y,z$ to eliminate the first component of $\vec{\Omega}$ (and also the last one since $\Omega$ is hermitian). The final $\tilde z$ transformation can be used to make the ${\Omega^+}'$ component real. 

An analogous conclusion can be reached concerning the elimination of $\Omega^+$ instead of $\Omega^{++}$.

\section{Different \texorpdfstring{$\boldsymbol{SU(2)'}$}{SU(2)'} Charge Assignments}
\label{app:reduciblerep}
Putting $\mu$, $e$ and $\tau$ in an $SU(2)'$-triplet seems natural, but is not the only possibility. We will now briefly discuss the other scheme, namely $e\sim \vec{1}$, $(\mu,\tau)\sim \vec{2}$, i.e.~the leptons form a reducible representation $\vec{1}\oplus \vec{2}$ under $SU(2)'$ (both left- and right-handed ones). Since the electron does not take part in the gauge interactions, there are no dangerous LFV involving the electron on tree-level, so a low $SU(2)'$ breaking scale is possible. The charged leptons now have masses $m_e$, $m_\mu = m_\tau$, so we need to break this symmetry using a Higgs field with leptospin-1. Putting the right-handed neutrinos in the same reps., i.e.~$N_e \sim \vec{1}$, $(N_\mu,N_\tau) \sim \vec{2}$ is problematic because only $N_e$ can acquire a Majorana mass term, the invariant $\epsilon_{a b} N^T_a N_b$ vanishes due to symmetry. We have therefore no good zeroth-order mass matrix, but would have to generate a proper $\mathcal{M}_R$ via $SU(2)'$ breaking.

Taking the right-handed neutrinos once again as a leptospin-1 field brings back the Majorana matrix~\eqref{eq:majoranamass}, but of course does not allow a Dirac mass term $\overline{\nu} \langle H \rangle N$, so with this assignment, $m_D$ has to be generated by $SU(2)'$ breaking. Both schemes provide bad starting points and seem unnatural, which is why we will not discuss them further.

It is, of course, possible to build viable models using different neutrino mass generation schemes than seesaw-I. In Refs.~\cite{horizontal_symmetry_reducible} the $\vec{1}\oplus\vec{2}$ representation was discussed in a similar context to build $L_e-L_\mu-L_\tau$ symmetric neutrino mass matrices, using either seesaw-II or $\mathcal{M}_R$ generation via VEVs, as discussed above.

\end{document}